\documentclass[dvips]{arxstspdf}
\usepackage{graphicx}
\usepackage{flushend}
\usepackage{stfloats}

\volume{23}
\issue{3}
\pubyear{2008}
\firstpage{332}
\lastpage{353}
\doi{10.1214/08-STS258}

\makeatletter

\newtheorem{prop}[thm]{Proposition}
\newtheorem{lem}[thm]{Lemma}

\newcommand{\Te}{\Theta}
\newcommand{\bet}{\beta}
\newcommand{\betk}{\beta_k}
\newcommand{\ei}{\varepsilon_i}
\newcommand{\sdue}{\sigma^2}
\newcommand{\sduek}{\sigma_k^2}

\newcommand{\cM}{\mathcal{M}}
\newcommand{\cMk}{\mathcal{M}_k}
\newcommand{\cIM}{\mathcal{IM}}

\newcommand{\ps}{\pi^{\mathrm{S}}}
\newcommand{\pkl}{\pi^{\mathrm{KL}}}
\newcommand{\puc}{\pi^{\mathrm{UC}}}

\newcommand{\Rex}{\mathbb{R}}

\def\epsilon{\varepsilon}

\newcommand{\sdueik}{\sigma_i^{2\,(k)}}
\newcommand{\muik}{\mu_i^{(k)}}
\newcommand{\yirep}{y_{i,\mathrm{rep}}}

\newcommand{\bkl}{b_k^{\mathrm{KL}}}
\newcommand{\dkl}{d_k^{\mathrm{KL}}}
\newcommand{\akl}{a_k^{\mathrm{KL}}}
\newcommand{\gkl}{g_k^{\mathrm{KL}}}
\newcommand{\betkl}{\bet_k^{\perp}}
\newcommand{\betkst}{\bet_k^{*}}
\newcommand{\sduekl}{{\sdue_k}^{\perp}}
\newcommand{\bnk}{b_{\setminus k}}

\makeatother

\begin{document}
\begin{frontmatter}

\title{Compatibility of Prior Specifications Across Linear Models}
\runtitle{Compatible priors for linear models}

\begin{aug}
\author[A]{\fnms{Guido} \snm{Consonni}\ead[label=e1]{guido.consonni@unipv.it}\corref{}}
and
\author[B]{\fnms{Piero} \snm{Veronese}\ead[label=e2]{piero.veronese@unibocconi.it}}
\runauthor{G. Consonni and P. Veronese}

\affiliation{University of Pavia and L. Bocconi University}

\address[A]{Guido Consonni is Professor,
Dipartimento di Economia Politica e Metodi Quantitativi,
University of Pavia, Via S. Felice 7, 27100 Pavia, Italy \printead{e1}.}
\address[B]{Piero Veronese is Professor, Department of
Decision Sciences, L. Bocconi University, Via Roentgen, 20136
Milano, Italy \printead{e2}.}

\end{aug}

%
\begin{abstract}
Bayesian model comparison requires the specification of a prior
distribution on the parameter space of each candidate model. In
this connection two concerns arise: on the one hand the
elicitation task rapidly becomes prohibitive as the number of
models increases; on the other hand numerous prior
specifications can only exacerbate the well-known sensitivity to
prior assignments, thus producing less dependable conclusions.
Within the subjective framework, both difficulties can be
counteracted by linking priors across models in order to achieve
simplification and compatibility; we discuss links with related
objective approaches. Given an encompassing,
or full, model together with a prior on its parameter space, we
review and summarize a few procedures for deriving priors under a
submodel, namely marginalization, conditioning, and
Kullback--Leibler projection. These techniques are illustrated and discussed
with reference to variable selection in linear models
adopting a conventional $g$-prior; comparisons
with existing standard approaches are provided. Finally, the relative merits
of each procedure are evaluated through simulated and real data
sets.
\end{abstract}

%
\begin{keyword}
\kwd{Bayes factor}
\kwd{compatible prior}
\kwd{conjugate prior}
\kwd{$g$-prior}
\kwd{hypothesis testing}
\kwd{Kullback--Leibler projection}
\kwd{nested model}
\kwd{variable selection}.
\end{keyword}

\pdfkeywords{Bayes factor, compatible prior, conjugate prior, $g$-prior, hypothesis testing, Kullback--Leibler projection, nested model, variable selection}

\end{frontmatter}

\section{Introduction}\label{sec:introduction}

Model comparison is an important and active area
of research especially from the Bayesian viewpoint; see, for example, George
(\citeyear{Geo1999}) and Robert (\citeyear{Rob2001}, Chapter 7). In
particular, the problem of
variable selection in linear models has received considerable
attention; see the review paper of George (\citeyear{Geo2000}) and a
few survey
chapters in the book edited by Dey and Rao (\citeyear{DeyRao2005}).
Two critical issues emerge from the very beginning: the
elicitation of prior probabilities for the various models under
consideration
and the assignment of prior distributions on the parameter
space of each model, which we simply call \emph{priors}. In this
paper we focus on the latter.

Occasionally, when the model space
is not large and detailed prior information is available, subjective
prior elicitation on each model can be carried out; see Garthwaite
and Dickey (\citeyear{GarDic1996}). More often, however, because of
the potentially
very high number of models under investigation, prior elicitation
can represent a formidable task, and hence practically implementable
procedures have been actively looked for. In the objective
framework (see Berger and Pericchi, \citeyear{BerPer1996b}), a
convenient approach is
to start with a default, typically improper, prior under each model,
and then to circumvent the indeterminacy of the normalizing constant
through an intrinsic prior procedure (see also Casella and Moreno,
\citeyear{CasMor2006}, for an application to variable selection in
linear models).
A more general approach, namely expected posterior prior,
is described in P\'erez and Berger (\citeyear{PerBer2002}).

Outside the purely objective view, pragmatic simplification of the
elicitation task in the variable selection problem has been achieved
through hierarchical mixture priors as in George and McCulloch
(\citeyear{GeoMcC1997}), or using an empirical Bayes approach, as in
George and Forster (\citeyear{GeoFos2000}), and more recently in Yuan
and Lin (\citeyear{YuaLin2005}), or employing a blend of noninformative
and conjugate procedures, as exemplified in Fern\'{a}ndez, Ley and
Steel (\citeyear{FerLeySte2001}). Recently Liang et al.
(\citeyear{LiaPauMolClyBer2008}) have proposed mixtures of $g$-priors
as an efficient tool for Bayesian variable selection.

Within the subjective framework, which uses proper priors,
the idea of relating priors across models does not seem to be
pervasive. Notable exceptions are Dickey (\citeyear{Dic1971}) and
Poirier (\citeyear{Poi1985}),
in the context of linear models; see also the discussion in O'Hagan
and Forster (\citeyear{OHaFor2004}, Sections 11.29--11.31). Neal
(\citeyear{Nea2001}) introduces the
idea of transferring prior information from a ``donor
model'' to a ``recipient model.'' His motivation
is primarily pragmatic: priors for complex models are harder to
elicit than those for simple models; accordingly one can try to carefully
elicit a prior under a simple ``donor'' model and then
transfer this information to a complex ``recipient''
model. Technically Neal's method is similar to, although more
general than, the expected posterior prior of Perez and Berger
(\citeyear{PerBer2002}). The paper by Dawid and Lauritzen (\citeyear
{DawLau2001}) stands out as an
attempt to discuss, in a general setting, methods to construct
``compatible priors'' for nested models using a variety of
strategies. Their motivation is mixed: on the one hand they state
that conceptually there is no compelling reason to relate priors
across models (since they express subjective opinions conditionally
on a different state of information); on the other hand such
relationships may be highly desirable on pragmatic grounds (the
effort spent in eliciting a prior under a model should somehow be
transferred to other models) and also to achieve some sort of compatibility
in order to lessen the sensitivity of the Bayes factor to prior specifications.

Following up this comment, we believe that priors for model comparison
deserve to be carefully investigated by
the Bayesian community. Traditional priors, which individually
perform quite effectively within a single model, need not work
satisfactorily when collectively employed for comparing models of
varying dimensions. This fact has been informally recognized at
least since Jeffreys, who refrained from using conventional priors
for comparing two nested hypotheses; see also Zellner and Siow
(\citeyear{ZelSio1980}) in the framework of linear models.

In the context of comparing a sharp null hypothesis $H_0$ versus a
composite alternative $H$, Morris (\citeyear{Mor1987}) argued forcibly
for the
prior under $H$ to be ``centered around $H_0$'';
otherwise the prior under $H$ would be ``wasting away'' prior
probability mass in regions that are often too
unlikely to be supported by the data, thus unduly favoring $H_0$,
as lucidly spelled out in Casella and Moreno (\citeyear{CasMor2007});
see also
Consonni and La Rocca (\citeyear{ConLaR2008}). Carefully extending
this argument
to several models would surely be of great value and interest in
order to enhance our understanding of the issue of compatibility
of priors for model comparison. While this paper falls short of
providing a comprehensive treatment of this point, it nevertheless
tries to offer some guidance for further reflection and research.
Specifically, we try to elucidate the meaning of the term ``submodel,''
or nested model, in order to highlight
differences between a couple of approaches which are implicit in
the literature and better understand specific strategies to relate
priors across models. Although the scope of our considerations is
general, we will illustrate the main ideas with reference to the
problem of variable selection in linear models.

The structure of the paper is as follows. Section
\ref{sec:submodels} deals with two notions of nested models and
discusses the corresponding parametrization, distinguishing
between nuisance and common parameters. Section
\ref{sec:strategies} deals with strategies to assign priors on
parameters of submodels starting from a prior on the (full) model;
we discuss conditioning and projection (including
marginalization) and propose, in Sections
\ref{sec:coherence nuisance} and \ref{sec:coherence nested},
two criteria to evaluate such
strategies, which we name nuisance- and nested-coherence.
Section \ref{sec:linear} deals with priors for linear models.
Starting with a $g$-prior under the full model, a variety of prior
specifications on submodels is obtained through the procedures
described in Section \ref{sec:strategies}; in particular Section
\ref{subsec:info-par} contains a discussion of the so-called
``information paradox.'' Section \ref{sec:examples}
presents three examples to evaluate the performance of the various
priors under consideration in terms of model comparison, with
special references to sensitivity issues. Finally, Section
\ref{sec:discussion} provides a few points for discussion. To ease
the flow of ideas, technical aspects have been relegated to the
\hyperref[app]{Appendix}.

\section{Submodels}\label{sec:submodels}

\subsection{A Preliminary Example}\label{subsec:a preliminary}

We start by discussing a very simple example with the aim of
presenting the main issues at stake. Consider the following model:
%
\begin{eqnarray}
\cM\dvtx\  y_i=\alpha+\beta x_i+\ei, \quad i=1,\ldots,n,
\nonumber\\
\eqntext{(\alpha,\beta,\sdue) \in\Te=\Rex\times\Rex\times\Rex^+}
\end{eqnarray}
where, conditionally on $\sdue$,
$\ei\stackrel{\mathrm{iid}}{\sim} \mbox{N}(0,\sigma^2)$.
An obvious submodel, say $\cM^*$, removes the predictor, thus changing
the mean structure.
However, several instances of $\cM^*$ are available, namely:
\begin{eqnarray*}
\cM^*_A\dvtx\  y_i&=&\alpha+\ei,
\quad\ei\stackrel{\mathrm{iid}}{\sim} \mbox{N}(0,\sigma^2);\\
\cM^*_B\dvtx\ y_i&=&\alpha+\ei^*,
\quad\ei\stackrel{\mathrm{iid}}{\sim} \mbox{N}(0,\sigma^{*2});\\
\cM^*_C\dvtx\  y_i&=&\alpha^*+\ei,
\quad\ei\stackrel{\mathrm{iid}}{\sim} \mbox{N}(0,\sigma^{2});\\
\cM^*_D\dvtx\ y_i&=&\alpha^*+\ei^*,
\quad\ei^* \stackrel{\mathrm{iid}}{\sim} \mbox{N}(0,\sigma^{*2}).
\end{eqnarray*}

Model $\cM^*_A$ originates in the setting of hypothesis testing
postulating that $\beta=0$ under $\cM$; in other words, $\cM^*_A$
is equivalent to the hypothesis $H^*\dvtx y \sim\cM$ and
$\beta=0$. As a consequence the parameters $\alpha$ and $\sdue$
are ``common'' to both models, although one might
further distinguish between them, since $\sdue$ pertains to the
error structure (which is not affected explicitly by the submodel
specification), and thus can be regarded as a ``nuisance''
parameter. Model $\cM^*_B$ originates from the consideration
that the error component in the submodel might, and perhaps
should, be allowed to be different from that under $\cM$. In
particular, since one can anticipate a worse fit under $\cM^*_B$
than under $\cM$, one should have $E(\sigma^{*2}) \geq E(\sdue)$
or even $\sigma^{*2} \geq\sdue$ (with probability 1). Model
$\cM^*_C$ originates from the consideration that the meaning of
the intercept is actually quite different under the two models,
and so should be distinct from that under $\cM$. On the other hand
$\sigma^2$ remains the same, since it is regarded as a ``nuisance''
parameter. Finally model $\cM^*_D$ combines the
specific features of $\cM^*_B$ and $\cM^*_C$, and has no direct
link, unlike the previous versions, to $\cM$. For a related
discussion on alternative interpretations of submodels, see
Berger and Pericchi (\citeyear{BerPer2001}, Section
1.5, ``Difficulty 4'').

In an abstract sense, all instances of $\cM^*$ above represent the
same submodel, since they share the same family of distributions.
However, the distinctive features that we have tried to underline
should make it clear that they are different objects, or perhaps
different ways of looking at the same object. For a given prior $\pi$
on $(\alpha,\beta, \sdue)$ under
$\cM$, we require a prior, $\pi^*$ say, under
$\cM^*$. We claim that each instance of $\cM^*$
naturally suggests a different procedure to obtain $\pi^*$ from~$\pi$.

Consider first model $\cM^*_A$. There are two natural candidates
for $\pi^*$, namely $\pi({\alpha,\sigma^2})$ and
$\pi({\alpha,\sigma^2|\beta=0})$, that is, the marginal and the
conditional (on $\beta=0$) distribution derived from
$\pi({\alpha,\beta,\sigma^2})$. The latter might appear more
natural, if the hypothesis-testing interpretation of $\cM^*_A$ is
strictly adhered to. Note that the two procedures lead to the
same priors if $(\alpha, \sigma^2)$ is independent of $\beta$, as
it occurs using default priors. For model $\cM^*_B$, instead, no
obvious indications are provided for the specification of
$\pi^*(\sigma^{*2})$; on the other hand, since $\alpha$ is
``common'' to both models, a natural suggestion would be to
take $\pi^*(\alpha)=\pi(\alpha)$. Of course the problem of
combining the two marginal distributions into a joint one remains
open. Under model $\cM^*_C$ a situation somewhat similar to that
under $\cM^*_B$ obtains, if we interchange the role of the
intercept and the variance. Finally, neither marginalization nor
conditioning appears as obvious recommendations under $\cM^*_D$,
because no effective link with $\cM$ is specified. The next
sections explore these issues in greater generality.

\subsection{Nested Models}\label{subssec:nested}

It could be argued that each of the models $\cM^*$ described in Section
\ref{sec:submodels} is
\emph{nested} in $\cM$. However, we feel some other clarification is needed.

Consider a model $\cM=\{f(\cdot|\theta), \theta\in\Te\}$. There
seem to be two interpretations of a nested model $\cM^*$ in the
literature, often not clearly distinguished. Both start from the
assumption that it is possible to write $\theta=(\lambda,\phi)$,
where $\lambda\in\Lambda$ and $\phi\in\Phi$, with $\lambda$
and $\phi$ being variation-independent, so that $\Te= \Lambda
\times\Phi$ and model $\cM^*$ is identified through the
constraint $\phi=\phi_0$, with $\phi_0$ a fixed value. As
suggested by a referee, this setting covers only the case in which
the parameter space $\Te^*$ associated with $\cM^*$ has
dimension strictly smaller than that of $\cM$, and thus it does
not account for other interesting nesting situations in which
dim$(\Te^*)=\operatorname{dim}(\Te)$ (e.g., when $\Te^*$ is a restriction of
$\Te$). However, the above $(\lambda, \phi)$-representation is
especially useful from the perspective of ``prior assignment'' under
submodels, which is the primary
focus of this paper. We describe these
two interpretations below.\\

S-N (\emph{Strongly nested interpretation}):
The sampling distribution of $y$ under $\cM^*$ is given by
$f^*(\cdot|\lambda), \lambda\in\Lambda$, where
$f^*(\cdot|\lambda)=f(\cdot|\lambda,\phi=\phi_0)$. This
interpretation can be clarified in terms of the underlying
generating process of $y$: ``If Nature chooses $\lambda
\in\Lambda$ and $\phi=\phi_0$, then the distribution of the
observables under $\cM$ and $\cM^*$ is the same.''\\

W-N (\emph{Weakly nested interpretation}):
The sampling distribution of the
observations $y$ under $\cM^*$ can be written as $
f^*(\cdot|\gamma), \quad\gamma\in\Lambda$, with $
f^*(\cdot|\gamma)=f(\cdot|\lambda=\gamma,\phi=\phi_0)$. In this
way $\gamma$, although structurally equivalent to $\lambda$, is
distinct from it. Clearly, each distribution in $\cM^*$ also
belongs to $\cM$.\\

Interpretation S-N is rooted in a hypothesis-testing context,
that is, $H^*\dvtx\phi=\phi_0$, where the actual objective of the
analysis is verifying whether $\phi=\phi_0$, \emph{other things
being held equal}. On the other hand, W-N is better suited when
the objective is model simplification, and each model competes
against the other ones according to whatever criterion is deemed
to be appropriate (e.g., a combination of fit and parsimony, or on
predictive grounds; see, e.g., Gelfand and Ghosh, \citeyear{GelGho1998} and
Marriott, Spencer and Pettitt, \citeyear{MarSpePet2001}). With regard
to the example in Section
\ref{subsec:a preliminary}, $\cM^*_A$ is the only instance of
$\cM^*$ that falls under interpretation S-N.
The \mbox{S-N} view is probably the most pervasive and is
regarded as a natural framework by, for example, Poirier (\citeyear
{Poi1985}), O'Hagan and
Forster (\citeyear{OHaFor2004}, Section
7.15) and Davison (\citeyear{Dav2003}, page 127). It seems implicit
in George and
Forster (\citeyear{GeoFos2000}) and other workers mostly interested in
computational aspects, for example, Smith and Kohn (\citeyear
{SmiKoh1996}), Nott and
Green (\citeyear{NotGre2004}) and Cripps, Carter and Kohn (\citeyear
{CriCar2005}). On the other hand,
authors like Berger and Periccchi
(\citeyear{BerPer1996a}) and also Robert (\citeyear{Rob2001}, Section~7.2) seem to prefer
interpretation W-N.

Within the interpretation S-N, consider a collection of submodels
$\cMk$ and suppose that, for each $\cM_k$, there exists a
reparametrization of $\cM$ as\break $(\delta, \eta_k, \omega_k)$, so
that $\cMk$ is identified by $\eta_k=\eta_{k0}$. Since $\delta$ is
never involved in any submodel specification we can regard it as a
\emph{nuisance} parameter; on the other hand we call $\omega_k$
the parameter \emph{common} to the pair $(\cM, \cMk)$.
In the setting of variable selection for linear models, the
nuisance parameter is clearly represented by the error variance
$\sdue$, while common parameters are the regression coefficients
that are not set to zero in the submodel specification.

We close this section with a caveat that hopefully will not disconcert
the reader.
Despite our insistence on model interpretation and parametric
description, we emphasize
that what matters in a Bayesian analysis is the \emph{prior distribution}
attached to the
parameters of the various models regardless of their formal representation.
The latter, however, may become relevant when structuring prior
specification across models.
This is the topic of the next section.

\section{Strategies to Assign Priors on Parameters of Submodels}\label{sec:strategies}

Within the objective Bayesian framework, the expected posterior prior
(EPP) methodology of P\'erez and
Berger (\citeyear{PerBer2002}) is a method to construct prior distributions
for model comparison; see also Neal (\citeyear{Nea2001}) for related concepts.
The idea is to start with a prior distribution under each model,
compute its posterior under ``imaginary''
observations, and formally average the posterior through a
marginal data distribution that is common to all models. The
method is quite general, but is especially effective if one starts
with a default, possibly improper, prior
under each model. In this way the EPP method allows to use
improper priors for model comparison through Bayes factors, or
posterior model probabilities, since the indeterminate normalizing
constants cancel out. More generally,
EPP is a method to make priors ``compatible'' across models,
through their dependence on a common marginal data distribution;
thus this methodology can be applied also with subjectively
specified (proper) prior distributions.

Although appealing and flexible, implementing the
EPP methodology may be problematic. First of all the choice of
the common distribution is not unique. For instance, there exist
at least two competing choices, namely that corresponding to the
``simplest'' model, if it exists, and that corresponding to the empirical
distribution, which requires the identification of a minimal
training sample; see Berger and
Pericchi (\citeyear{BerPer2004}) for a discussion of potential
difficulties associated
to this concept.
More importantly, to judge the relative merits of the above two
choices is not straightforward. A second concern refers to the actual
implementation of the EPP, which may require careful
computational strategies.

A more specific approach is the intrinsic prior methodology, which
has received a great deal of attention both for hypothesis testing
and for model selection. Again the primary motivation is the use of
default noninformative priors under each model; see Pericchi
(\citeyear{Per2005}) for a review. When several models are entertained the
intrinsic method requires a nesting strategy. One approach,
labeled ``encompassing from above,'' chooses as
benchmark a full model wherein all other models are nested. In
this way, however, the prior under the full model changes in each
pairwise comparison, thus producing an overall incoherent
probabilistic answer. Yet posterior probabilities can still be
formally defined on the basis of the collection of Bayes factors
of each model relative to the full one; see Casella and Moreno
(\citeyear{CasMor2006}) for an application to variable selection in
linear models.
On the other hand, if the simplest model (i.e., one being nested
within any other model) is available, an alternative ``encompassing
from below'' intrinsic prior procedure can be
followed, which is probabilistically correct; for an application
to variable selection see Moreno and Giron (\citeyear{MorGir2007}).
Notice that the
two alternative encompassing procedures will typically lead to
distinct answers. As with the EPP methodology, analytic evaluation
of intrinsic priors is typically very hard and actual
implementation of the procedure requires a good deal of
computational ingenuity; see Casella and Moreno (\citeyear
{CasMor2005}) in the
context of contingency tables.

Although the EPP and intrinsic prior methodologies produce priors
that are ``related'' through a common underlying
marginal data distribution, they do not explicitly address the issue of
prior compatibility across models. The latter issue is lucidly tackled
in Dawid and Lauritzen (\citeyear{DawLau2001}), who present several
strategies for the derivation of compatible priors; see also
Roverato and Consonni (\citeyear{RovCon2004}) in the context
of directed graphical models and Consonni, Guti\'errez-Pe\~na and
Veronese (\citeyear{ConGutVer2007})
for general exponential families with a detailed application to
testing the Hardy--Weinberg model in studies of population
genetics.

Starting with a model $\cM=\{ f(y|\lambda,\phi) \}$ and a joint
distribution $\pi({\lambda,\phi})$, we briefly review below four
main strategies for prior specification under a nested model
$\cM^*$ identified through $\phi=\phi_0$.

\emph{Marginalization} (M). This approach is most natural under
interpretation S-N where $\cM^*=\{f^*(y|\lambda),\break \lambda\in
\Lambda\}$, so that $\cM$ and $\cM^*$ share the same parameter
$\lambda$, and states that $ \pi^{\mathrm{M}}({\lambda})=\pi({\lambda})$,
where $\pi({\lambda})$ is the marginal of $\lambda$ under
$\pi({\lambda,\phi})$.
Two critical aspects should be taken into consideration:
(i) marginalization does not explicitly take into consideration the constraint
$\phi=\phi_0$; in fact it disregards this information by averaging
with respect to the distribution of $\phi$;
(ii) on a
more technical side, this procedure is not invariant to
reparametrization. Consider, for instance, model $\cM$ of Section~\ref{subsec:a preliminary}, and suppose to recenter the data as
$x_i \rightarrow x_i - \bar{x}$, with $\bar{x}$ the mean of the
$x_i$. The model~$\cM$ becomes $(\alpha-\beta\bar{x})+\beta x_i$
suggesting the following reparametrization: $(\alpha,\beta)
\mapsto(\gamma,\delta)$, where $\gamma=\alpha-\beta\bar{x}$,
and $\delta=\beta$. Notice that $\alpha$ and $\gamma$ are the same
quantities under $\cM^*$ and so should share the same prior under
the latter model. On the other hand, $\alpha$ and $\gamma$ are
distinct under $\cM$ and will have typically different priors, a
feature which will be inherited under $\cM^*$ through the
procedure M, thus establishing its lack of invariance.

\emph{Usual conditioning} (UC). As with M, this procedure applies
more naturally under interpretation S-N, and states that $
\pi^{\mathrm{UC}}({\lambda})=\pi({\lambda|\phi=\phi_0})$,
where the right-hand side is the conditional distribution of
$\lambda$ given $\phi=\phi_0$ under $\pi(\lambda,\phi)$. A clear
advantage of UC is that it incorporates explicitly the information
available in the specification of model $\cM^*$, through the
constraint $\phi=\phi_0$. The major drawback of UC is that it is
not invariant to the choice of the conditioning function
(typically an event having zero probability) which identifies the
submodel. For instance, assume that $\cM$ is as in Section
\ref{subsec:a preliminary}, and that $(\alpha, \beta)$ are
jointly normal with zero mean, variances $\sdue_{\alpha}$,
$\sdue_{\beta}$ and correlation coefficient~$\rho$. Then the
distribution of $\alpha$ given $\beta=0$ is normal with zero mean
and variance $\sdue_{\alpha}(1-\rho^2)$. On the other hand, model
$\cM^*$ could also be identified through the constraint $\xi=0$,
where $\xi=\beta/\alpha$. It can be checked that the conditional
distribution of $\alpha$ given $\xi=0$ is no longer normal. This
represents an instance of the Borel--Kolmogoroff paradox.

\emph{Jeffreys conditioning} (JC). This procedure is a variation
of UC and hence is most appropriate again under interpretation
S-N. It was proposed by Dawid and Lauritzen (\citeyear{DawLau2001}) to overcome
the lack of invariance of UC. First recall that the density
obtained through UC can be expressed as
$\pi^{\mathrm{UC}}(\theta) \propto\pi(\theta), \quad\theta\in\tilde{\Te}^*$,
where $\tilde{\Te}^*=\{(\lambda,\phi), \lambda\in\Lambda,
\phi=\phi_0 \}$. Now let $H(\theta)$ denote the Fisher information
matrix for $\theta$ under $\cM$, and similarly for $H^*(\theta)$
under $\cM^*$. Set $j(\theta) \propto|H(\theta)|^{1/2}$, where
$|H|$ is the determinant of $H$, so that $j(\theta)$ is the
Jeffreys prior for $\theta$ under $\cM$, and define analogously
$j^*(\theta)$ under model $\cM^*$. The JC density is defined as
%
\begin{equation}\label{A:jeffreysconditioning}
\pi^{\mathrm{JC}}(\theta)\propto\pi(\theta) \frac{j^*(\theta)}{j(\theta
)}, \quad\theta
\in\tilde{\Te}^*.
\end{equation}
Typically,
one would re-express the JC density as a function of $\lambda$ only,
and write
$\pi^{\mathrm{JC}}({\lambda})$ accordingly; we shall follow this style in the
next section.
A useful feature of Jeffreys conditioning is invariance to
model reparametri\-zation, because of the multiplicative term given by
the ratio of the Jeffreys
densities. A potential difficulty with Jeffreys conditioning
is that the resulting prior $\pi^{\mathrm{JC}}({\lambda})$ may be improper
even though $\pi({\theta})$ is proper, because of its
nonprobabilistic nature.

\emph{Kullback--Leibler \textup{(}KL\textup{)} projection}. This procedure is part of
a more general approach to the construction of priors on related
models based on \emph{projection maps}, and is especially
appropriate under interpretation \mbox{W-N}.
Consider a model $\cM$ and a submodel $\cM^*$, parametrized
by $\theta^* \in\Te^*$ for the same observable, and suppose that each
distribution in
$\cM$ has an image in $\cM^*$ through the (projection) map $\tau
\dvtx
\Te\mapsto\Te^*$. Given a prior $\pi(\theta)$ on $\Te$, the prior
induced on $\tau(\theta)$ is called the $\tau$-projection prior.

For reasons to be specified shortly below, we shall take $\tau(\theta)$
as the Kullback--Leibler (KL)-projection of $\theta$ onto $\Te^*$,
that is,
\[
\tau_{\theta}^{\mathrm{KL}}(\theta)= \arg\min_{\theta^* \in\Te^*}
\mathit{KL}(f(\cdot|\theta),f^*(\cdot|\theta^*)) ,
\]
where
\[
\mathit{KL}(p,q)=E^p \biggl(\log\frac{p(X)}{q(X)} \biggr)
\]
denotes the KL-divergence between the density $p$ and $q$ relative
to a common dominating measure. In this case we call the resulting
prior KL-projection prior, or KL-prior for short, and denote it
with\break $\pi^{\mathrm{KL}}({\theta^*})$, that is, $
\pi^{\mathrm{KL}}({\theta^*})=\pi^{\theta}_{\theta^{\perp}}(\theta^*)$,
where
$\pi^{\theta}_{\theta^{\perp}}$ is the prior on
$\theta^{\perp}=\tau^{\mathrm{KL}}_{\theta}(\theta)$ induced from the prior
$\pi({\theta})$. KL-priors were originally presented in McCulloch and
Rossi (\citeyear{McCRos1992}) to compute Bayes factors; they are
applied in Viele
and Srinivasan (\citeyear{VieSri2000}) to ANOVA models, and in
Consonni, Guti\'errez-Pe\~na and
Veronese (\citeyear{ConGutVer2007}) to a particular multinomial\break model.
Goutis and Robert
(\citeyear{GouRob1998}) and Dupuis and Robert (\citeyear{DupRob2003})
use KL-projection for
comparing models, but do not rely on the idea of KL-priors.

Notice that $\mathit{KL}(p, q)$ is not symmetric. The intrinsic discrepancy
between $p$ and
$q$, $\delta(p,q)=\min\{ \mathit{KL}(p,\break q),\mathit{KL}(q,p)\}$ (see Bernardo and
Rueda, \citeyear{BerRue2002}), overcomes this difficulty. However, we
will still use
$\mathit{KL}(p,\break q)$ because (i) we
take $p$ as the encompassing model, whose validity is not
questioned within our approach, while $q$ is a simplified version of
$p$; from this
point of view taking expectations with respect to $p$, as in
$\mathit{KL}(p,q)$, appears a sensible procedure; (ii) for regular nested
models (wherein the support is independent of the parameter), $p$
and $q$ have the same support so that $\mathit{KL}(p,q)$ is well defined;
(iii) the use of $\delta(p,q)$, instead of $\mathit{KL}(p,q)$, adds
complexity from an analytical viewpoint (for a detailed discussion
on these points see Consonni, Guti\'errez-Pe\~na and
Veronese, \citeyear{ConGutVer2007}).

From our perspective, a very important feature of the
KL-projection is its invariance to reparametrization. Thus if
$\eta=g(\theta)$ is a reparametrization under $\cM$, then
$\tau_{\eta}^{\mathrm{KL}}(\eta) = \tau_{\theta}^{\mathrm{KL}}(g^{-1}(\eta))$.
Accordingly, prior assignments based on KL-projection do not
depend on the specific parametrization that is chosen.
To illustrate the KL-procedure, consider the simple linear model $\cM$ of
Section \ref{subsec:a preliminary} with the submodel specified
by $\cM^*_D$.
It can be checked that the KL-projection of $(\alpha,\beta,\sdue)$
onto the space $\{ (\alpha^*,\sigma^{*2}) \in\Rex\times\Rex^+ \}$
is given by
\begin{eqnarray*}
(\alpha,\beta,\sdue)^{\perp}&=&\biggl(\alpha+\beta\bar{x},
\sdue+\beta
^2\frac{1}{n}\sum(x_i-\bar{x})^2\biggr)\\
&=&(\alpha^{\perp}, \sigma^{2\perp}),
\end{eqnarray*}
with some abuse of notation for the latter equality.
It is interesting to remark that the projection corresponding to the variance
is given by $\sdue$ plus a quadratic term: as a consequence
$\sigma^{*2 }$ is stochastically larger, under the KL-prior, than
$\sdue$, whatever the prior on $\sigma^{2}$ under $\cM$. This seems
to be consistent with the views of those authors who state that
$\sigma^{*2}$ should perhaps be larger than $\sdue$, to account
for an anticipated worse fit of the submodel; see Berger and
Pericchi (\citeyear{BerPer2001}, Section 1.5) and Robert (\citeyear
{Rob2001}, page 349). A similar,
although less stringent, view is held by George and McCulloch
(\citeyear{GeoMcC1997}) according to whom the \emph{expectation} of
$\sdue$ under
the smaller model should be larger.
The exact form of the joint KL-prior for $(\alpha^*,\sigma^{*2})$
is typically unavailable because of the complicated structure of
$\sigma^{2 \perp}$; however, we will provide an analytical
approximation in the next section. Alternatively, one could resort to
stochastic simulation since a draw from $\pi^{\mathrm{KL}}(\cdot)$ can be
easily obtained by first generating $\tilde{\theta}$ from $\pi(\cdot
)$ and then
calculating $\tau_{\theta}^{\mathrm{KL}}(\tilde{\theta})$, possibly through numerical
methods.

\subsection{Coherence of Procedures With Respect to Nuisance
Parameters} \label{sec:coherence nuisance}

In this section we plan to evaluate the procedures to construct
priors under submodels from the point of view of coherence with
respect to the nuisance parameter as defined in Section
\ref{subssec:nested}.

If $\delta$ is a nuisance parameter, then it could be integrated out from
the very beginning (see O'Hagan and Forster, \citeyear{OHaFor2004},
Sections 3.13--3.14), using a prior under $\cM$. A new \emph{integrated}
model $\mathcal{IM}$ would then be obtained, which in turn generates an
integrated submodel $\mathcal{IM}^*$.
Let $y$ be a future observation to be forecast. We say that
a procedure is \emph{nuisance-coherent} if the marginal
distributions of $y$ under submodel $\cM^*$ and the
corresponding integrated submodel $\cIM^*$ are the same, that is,
%
\begin{equation}\label{eq:nuisance coherence}
f^*_{\cM^*}(y)=f^*_{\cIM^*}(y).
\end{equation}
In other words, integrating out the nuisance parameter ``at
the beginning'' (using $\pi$) or ``at the end''
(using the procedure-induced prior) does not make any difference.
If (\ref{eq:nuisance coherence}) holds, then the predictive
distributions under the two models are equivalent; moreover, the
Bayes factor for the pair $(\cM,\cM^*)$
coincides with that for $(\cIM,\cIM^*)$, since
$f_{\cM}(y)=f_{\cIM}(y)$ by definition of integrated model.

The following proposition establishes results on
nuisance-coherence for the procedures M, UC and JC.

\begin{prop}
\label{prop:nuisance coherence}
Consider a model $\cM$ para\-me\-trized by $(\lambda,\delta,\phi)$ with
$\delta$ a nuisance
parameter, and prior $\pi(\lambda,\delta,\phi)$.
Let $\cM^*$ be a submodel identified\break through $\phi=\phi_0$.
Then:
\begin{longlist}[(iii)]
\item[(i)]
the UC procedure is nuisance-coherent;
%
\item[(ii)]
the M procedure is nuisance-coherent if $\delta$ is conditionally
independent of $\phi$ given $\lambda$ under
$\pi(\lambda,\delta,\break \phi)$;
\item[(iii)]
the JC procedure is nuisance-coherent if the ratio of the
Jeffreys priors relative to the pair $(\cM,\cM^*)$ is
proportional to that for the pair $(\cIM,\cIM^*)$, provided the
resulting priors are proper.
\end{longlist}
\end{prop}
\begin{pf}
See the \hyperref[app]{Appendix}.
\end{pf}

In general nuisance-coherence does not hold for the KL-procedure; see
Section \ref{subsubsec:kullback}.

\subsection{Coherence of Procedures Across Nested Models}
\label{sec:coherence nested}

We now address the issue of coherence
across a collection of submodels. It is actually enough to
consider only three models. For simplicity of exposition we shall
formulate the problem within interpretation S-N (see Section \ref
{subssec:nested}).
Specifically, consider the following models:
%
\begin{eqnarray}\label{eq:nested coherence1}
&&\quad\cM\dvtx\phantom{**} \  f(y|\lambda,\phi_1,\phi_2),\\
&&\quad\cM^* \dvtx\phantom{*} \ f^*(y|\lambda,\phi
_2)=f(y|\lambda,\phi_1=\phi
_1^0,\phi_2), \label{eq:nested coherence2}\\
\label{eq:nested coherence3}
&&\quad\cM^{**} \dvtx\
f^{**}(y|\lambda)=f(y|\lambda,\phi_1=\phi_1^0,\phi_2=\phi
_2^0)\nonumber\\[-8pt]\\[-8pt]
&&\ \phantom{\cM^{**} \dvtx\quad
f^{**}(y|\lambda)}
=f^*(y|\lambda,\phi_2=\phi_2^0),\nonumber
\end{eqnarray}
so that $\cM^*$ is a submodel
of $\cM$ and $\cM^{**}$ is a submodel of $\cM^*$ (and so also of
$\cM$). Let $\pi(\lambda,\phi_1,\phi_2)$ be the prior under $\cM$,
$\pi^*(\lambda,\phi_2)$ that under $\cM^*$ and finally
$\pi^{**}(\lambda)$ that under $\cM^{**}$. For each given
procedure to construct priors on submodels, the prior
$\pi^{**}(\lambda)$ can be obtained either with respect to the
pair $(\cM,\cM^{**})$, which we label $\pi^{**}_{\cM}(\lambda)$,
or with respect to the pair $(\cM^*, \cM^{**})$, which we label
$\pi^{**}_{\cM^*}(\lambda)$.

We say that a procedure is \emph{nested-coherent} if\break $
\pi^{**}_{\cM}(\lambda)=\pi^{**}_{\cM^*}(\lambda)$.

\begin{prop} \label{prop:nested coherence}
Consider the three models described in \textup{(\ref{eq:nested
coherence1})--(\ref{eq:nested coherence3})}.
The M, UC and JC procedures are nested-coherent.
\end{prop}

\begin{pf}
See the \hyperref[app]{Appendix}.
\end{pf}

We remark that nested-coherence fails in general for the KL-procedure
as we report in Section \ref{subsubsec:kullback} with
reference to linear models.

\section{Linear Models}\label{sec:linear}

Consider the general linear model $\cM$
%
\begin{equation}\label{eq:general linear model}
y=X\bet+ \epsilon,
\end{equation}
where $y$ is an $n$-dimensional vector of observations on the dependent
variable,
$X$ an $(n\times p)$ matrix of predictors having rank $p$, $\beta$ a
$p$-dimensional vector
of regression coefficients and $\epsilon$ an $n$-dimensional vector of
error terms with
$\epsilon\sim\mbox{N}(0,\sdue I)$, conditionally on $\sdue$.
We assume that the constant term
is always included in the model, so that the first column of $X$ is
the unit vector.
It is useful to think of (\ref{eq:general linear model}) as the
\emph{full} model.

If subjective information is limited,
we can easily resort to conventional proper priors such as the
conjugate normal inverted gamma (NIGa) family; see, for example, O'Hagan
and Forster (\citeyear{OHaFor2004}, Section 11.4).
Specifically, under a NIGa$(b,V,d,a)$ prior, the conditional
distribution of $\bet$ given $\sdue$ is N$(b,\sdue V)$ while the
marginal distribution of $\sdue$ is IGa$(d/2,a/2)$. Here,
N$(b,\Sigma)$ denotes a normal distribution with expectation $b$
and variance matrix $\Sigma$, while IGa$(d/2,a/2)$ stands for an
inverted gamma distribution having expectation $a/(d-2), d>2$.
In many applications, and especially in econometric analysis, a
simplified version of the NIGa prior is usually considered. The
suggestion of Zellner (\citeyear{Zel1986}), called \emph{$g$-prior},
is to set
$V=g(X^TX)^{-1}$, with $g>0$. The choice of $g$ has been
extensively analyzed in several papers, for example, George and Foster
(\citeyear{GeoFos2000}), Clyde and George (\citeyear{ClyGeo2004}) and
Fern\'{a}ndez, Ley and Steel
(\citeyear{FerLeySte2001}).

Some authors have raised criticism against the use of $g$-priors
for model selection (see for a clear exposition Berger and
Pericchi, \citeyear{BerPer2001}), and have suggested alternative
conventional priors, such
as the Cauchy prior by Zellner and Siow (\citeyear{ZelSio1980}), recently
discussed in Bayarri and Garcia-Donato (\citeyear{BayGar2007}). Liang
et al.
(\citeyear{LiaPauMolClyBer2008}) propose to use a prior on the
parameter $g$ leading to a
mixture of $g$-priors, which includes as a special case that by
Zellner and Siow. This prior does not suffer from the ``information
paradox'' which represents a major drawback of
$g$-priors; see Section \ref{subsec:info-par}.
However, we still \mbox{employ} a $g$-prior on the full model because of its simplicity
and analytical tractability. At any rate the
compatible priors that we derive under the various submodels differ
from the
$g$-priors traditionally employed.

We take as prior for $(\bet, \sdue)$ under $\cM$
%
\begin{eqnarray}\label{eq:gNIGa M}
\pi(\bet,\sdue)&=&
\mbox{NIGa}(\bet,\sdue;b,g(X^TX)^{-1},d,a) ,
\end{eqnarray}
hierarchically specified through
\begin{eqnarray}\label{eq:g-prior M}
\pi(\bet|\sdue)&=& \mbox{N}(\bet; b,g
\sdue(X^T X)^{-1}); \nonumber\\[-8pt]\\[-8pt]
\pi(\sdue) &=& \mbox{IGa}(\sdue; d/2,a/2),\nonumber
\end{eqnarray}
and refer informally to (\ref{eq:gNIGa M}) as the gNIGa
prior.

Concerning the choice of $E(\bet)=b$, three default options are
%
\begin{eqnarray}\label{eq:m-default}
\qquad b^T_0&=&(0,\ldots,0), \quad{\bar b}^T=(\bar{y},0,\ldots,0),
\quad{\hat b}=\hat{\bet},
\end{eqnarray}
where $\hat{\bet}$
represents the OLS estimate of $\bet$ under the full model. In
this way the elicitation of the gNIGa prior reduces simply to
choosing the three hyperparameters $d,a$ and $g$.
Possible choices for $g$ are extensively discussed in
Fern\'{a}ndez, Ley and Steel (\citeyear{FerLeySte2001}). In
particular, based on
simulation results, they recommend using $g=\max\{ n,p^2
\}$, so that typically $g=n$, because $n$
ordinarily exceeds $p^2$.

\subsection{Priors for Submodels}\label{subsec:priors}
We now review some techniques
for prior specification under a generic linear submodel.
Let $\cM_k$ represent a submodel that uses $p_k$ predictors with\vspace*{-5pt}
$p_k <p$. Write $X=(X_k \vdots X_{\backslash k})$,
where $X_k$ is an $(n\times p_k)$ matrix.
We assume that each submodel includes the intercept term, so that the
first column of $X_k$ is the unit vector;
for this reason there exist $2^{p-1}$ possible models. Let
$\bet^T=(\bet_k^T,\bet_{\backslash k}^T)$ be the partition
corresponding to that of $X$.

If we adopt interpretation S-N of nested models, we can write $\cM_k$ as
$y=X_k \bet_k + \epsilon$,
which is equivalent to the hypothesis $H_k\dvtx\bet_{\backslash
k}=0 $. On the other hand if one follows interpretation W-N,
$\cM_k$ can be expressed as
%
\begin{eqnarray}\label{eq:submodel mod selection}
y=X_k\bet^*_k + \epsilon_k,
\end{eqnarray}
with $\epsilon_k \sim\mbox{N}(0,\sdue_k I)$,
and $\bet^*_k$ a $p_k$-dimensional vector. Notice that in this
setting each submodel presents a specific parametric
representation, with a distinct $\bet_k^*$ and $\sdue_k$. To
simplify the exposition, in the following we will make use
exclusively of representation (\ref{eq:submodel mod selection})
which reduces to the S-N case by setting $\bet^*_k=\bet_k$ and
$\sdue_k=\sdue$.

It is common practice to ``replicate'' the gNIGa
prior described in (\ref{eq:gNIGa M}), under each $\cM_k$, in
particular using the same values of $g$, $d$ and $a$. We will show
that the UC and JC procedures, as well as KL based on a conjugate
approximation, lead instead to
\begin{eqnarray}\label{eq:gNIGa Mk}
&&\pi_k(\bet^*_k,\sdue_k)\nonumber\\[-8pt]\\[-8pt]
&&\quad=
\mbox{NIGa}(\bet^*_k,\sdue
_k;b^*_k,g_k(X_k^TX_k)^{-1},d_k,a_k),\nonumber
\end{eqnarray}
with model-specific hyperparameters. As a consequence,
the marginal distribution of $y$ is an $n$-dimensio\-nal Student
t-distribution and the Bayes factor for model $\cM_k$ versus
model $\cM_s$ can be written as
%
\begin{eqnarray}\label{eq:Bks}
B_{ks}&=&C_{ks}\biggl\{a_s+\frac{g_s}{1+g_s}y^TM_sy\nonumber\\
&&\phantom{C_{ks}\biggl\{}
{}+\frac{1}{1+g_s}(y-X_sb^*_s)^T\nonumber\\
&&\hspace*{59pt}
{}\cdot(y-X_s b^*_s) \biggr\}^{(d_s+n)/2}\nonumber\\[-8pt]\\[-8pt]
&&{}\cdot
\biggl[\biggl\{a_k+\frac{g_k}{1+g_k}y^TM_ky\nonumber\\
&&\phantom{{}\cdot
\biggl[\biggl\{}
{}+\frac{1}{1+g_k}(y-X_kb^*_k)^T\nonumber\\
&&\hspace*{61pt}
{}\cdot
(y-X_k b^*_k) \biggr\}^{(d_k+n)/2}\biggr]^{-1};\nonumber
\end{eqnarray}
where
\begin{eqnarray*}C_{ks}=
\frac{\Gamma(d_s/2)(a_k)^{d_k/2}\Gamma((d_k+n)/2)(1+g_s)^{p_s/2}}
{\Gamma(d_k/2)(a_s)^{d_s/2}\Gamma((d_s+n)/2)(1+g_k)^{p_k/2}},
\end{eqnarray*}
with $M_k=I-X_k(X_k^TX_k)^{-1}X_k^T=I-P_k$, where $P_k$ is the
projection matrix
onto the column space of $X_k$.
Accordingly $y^TM_ky$ represents the residual sum of squares $\mathit{SSR}_k$ of model
$\cM_k$ and similarly for $M_s$.

Notice that the marginalization procedure does not lead to the gNIGa
prior (\ref{eq:gNIGa Mk}). Indeed,
conditionally on $\sdue_k$, the variance matrix of
$\bet_k^*$ is given by $g \sdue_k[(X^T\cdot X)^{-1}]_{kk}$, where
$[(X^TX)^{-1}]_{kk}$ is the submatrix of $(X^TX)^{-1}$ containing
the first $k$ rows and $k$ columns, which is not equal to $(X_k^TX_k)^{-1}$.
This reason, together with the lack of invariance and of
nuisance-coherence of the marginalization procedure in this case,
suggest to disregard it in
our future investigations.

\subsubsection{\texorpdfstring{Standard Approach.}{Standard Approach}}\label{subsubsec:standard}

The conventional prior that is used in most Bayesian analyses of linear models
\emph{assumes} that, under $\cM_k$, $(\bet^*_k,\sdue_k)$ follows a
gNIGa distribution,
with hyperparameters $(b^{\mathrm{S}}_k,g,d,a)$, where the
superscript $S$ stands for ``standard.''
Often the prior on
$\sdue_k$ is taken to be improper ($d \rightarrow0$ and $a
\rightarrow0$) and the resulting prior will be
denoted with $\pi^{\mathrm{I}}(\beta^*_k,\sduek)$,
where $I$ stands for ``improper.'' Standard choices
for $b^{\mathrm{S}}_k$ reproduce the default options (\ref{eq:m-default})
and can be formally recovered as $b_k^{\mathrm{S}}=(X_k^T X_k)^{-1}X_k^T Xb$.
Using results in Rao and Toutemburg (\citeyear{RaoTou1999}, pages
41--42), it can be
checked that when $b=\hat{b}$ the corresponding $b_k^{\mathrm{S}}$ will
coincide with the OLS estimate of $\beta_k$ under $\cM_k$.

We conclude this section remarking that the standard approach does not satisfy
nuisance-coherence (it is enough to check that the marginal
variance of $y$ under $\cM_k$ differs from that under $\cIM_k$);
on the other hand nested-coherence trivially holds.

\subsubsection{\texorpdfstring{Usual Conditioning.}{Usual Conditioning}}\label{subsubsec:usual}

The prior for $(\bet^*_k, \sdue_k)$ in this
case is given by
%
\begin{eqnarray}\label{eq:UC}
\pi_k^{\mathrm{UC}}(\bet^*_k,\sdue_k)&=& \pi(\bet^*_k, \sdue_k|\bet
_{\backslash k}=0)\nonumber\\
&=& \pi(\bet^*_k|\bet_{\backslash k}=0, \sdue_k)\pi(\sdue_k|\bet
_{\backslash k}=0) \\
& = &\pi^{\mathrm{UC}}_k(\bet^*_k|\sdue_k)\pi_k^{\mathrm{UC}}(\sdue_k).\nonumber
\end{eqnarray}

It can be checked that the UC prior is gNIGa, that is,
%
\begin{eqnarray}
\label{eq:gNIGa UC}
&&\pi_k^{\mathrm{UC}}(\bet^*_k,\sdue_k)\nonumber\\
&&\quad=
\mbox{NIGa}(\bet^*_k,\sdue
_k;b^{\mathrm{UC}}_k,\\
&&\phantom{\quad=\mbox
{NIGa}(}g_k^{\mathrm{UC}}(X_k^TX_k)^{-1},d^{\mathrm{UC}}_k,a^{\mathrm{UC}}_k)\nonumber
\end{eqnarray}
with
%
\begin{eqnarray}\label{dkUC-1}
b_k^{\mathrm{UC}}&=&b_k+(X_k^T X_k)^{-1}(X_{k}^T X_{\setminus k}) b_{\setminus
k},\nonumber\\[-8pt]\\[-8pt]
g^{\mathrm{UC}}_k&=&g, \quad d_k^{\mathrm{UC}}= d+ (p-p_k), \nonumber\\
\label{dkUC-2}
a_k^{\mathrm{UC}}&=& a+b_{\setminus k}^T X_{\setminus
k}^TM_kX_{\setminus k}b_{\setminus k}.
\end{eqnarray}
Analogous results were
derived in Poirier (\citeyear{Poi1985}). Notice that under UC the
hyperparameters change across models. In particular $d_k^{\mathrm{UC}}$
increases as $p_k$ decreases (the model becomes smaller).
George and McCulloch (\citeyear{GeoMcC1997}) also allow different
priors for the
variance under the various models, although their choice is not
based on formal probabilistic derivations. In their case, the
larger the model, the smaller the expected variance, which is not
necessarily the case under UC.
Notice that if $b_{\setminus k}=0$,
one obtains $E(\sdue_k)=a/(d_k^{\mathrm{UC}}-2)$, which
decreases as $p_k$ decreases. While this feature
may appear somewhat counterintuitive, it will turn
out to have useful implications as detailed in Section \ref{subsec:info-par}.

\subsubsection{\texorpdfstring{Kullback--Leibler Projection.}{Kullback--Leibler Projection}}\label{subsubsec:kullback}

The following lemma is instrumental in deriving KL-projections.
\begin{lem}\label{lemma}
Consider the linear model $\cM$ defined in (\ref{eq:general linear model}),
and the submodel $\cM_k$ defined in (\ref{eq:submodel mod
selection}). Then
\begin{longlist}[(iii)]
\item[(i)] the KL-divergence between $\cM$ and $\cM_k$ is given by
\begin{eqnarray*}
\mathit{KL}(\cM,\cM_k)&=&\frac{1}{2\sdue
_k}(X \bet-X_k
\bet^*_k)^T(X
\bet-X_k \bet^*_k)\\
&&{}+ \frac{n}{2 } \biggl[\frac{\sdue}{\sdue_k}-\log
\biggl(\frac{\sdue}{\sdue_k} \biggr)-1 \biggr];
\end{eqnarray*}
\item[(ii)]
%
\begin{eqnarray}\label{eq:KL-projection-betak}
\ \qquad\arg\min_{\beta_k^*}\mathit{KL}(\cM,\cM_k)=\beta
_k^{\perp}=(X^T_k
X_k)^{-1}X_k^TX\beta;\hspace*{-5pt}
\end{eqnarray}
\item[(iii)]
$\arg\min_{\beta_k^*,\sduek}\mathit{KL}(\cM
,\cM_k)
=(\beta_k^{\perp}, \sigma_k^{2 \perp})$,
\end{longlist}
where $\beta_k^{\perp}$ is defined in
(\ref{eq:KL-projection-betak}) and
%
\begin{equation}\label{eq:KL-projection-sduek}
\sigma_k^{2\perp}=\sigma^2+Q_k(\bet),
\end{equation}
with
\begin{eqnarray}
Q_k(\bet)&=&\frac{1}{n}\bet^TX^T M_k X\beta\nonumber\\[-8pt]\\[-8pt]
&=&\frac{1}{n}\bet^T_{\setminus k}X_{\setminus k}^TM_k X_{\setminus
k}\bet_{\setminus k}.\nonumber
\end{eqnarray}
\end{lem}

\begin{pf}
Point (i) follows specializing to our case the
KL-divergence between two multivariate normal distributions, given
for example in Whittaker (\citeyear{Whi1990}, page 387). Points (ii)
and (iii) are
obtained by a direct calculation.
\end{pf}

We now distinguish two cases, namely projection with
respect to $\bet_k^*$ for given $\sdue_k$, and projection
with respect to both $\bet_k^*$ and $\sdue_k$.
Consider the former case. This is appropriate, for instance, when we
want to
take the same prior on $\sdue_k$ for all models; in this case we
need only minimize $\mathit{KL}(\cM,\cM_k)$ with respect to $\beta_k^*$ and
thus $\beta_k^{\perp}$ is given by (\ref{eq:KL-projection-betak})
(for interesting related results, obtained using a predictive
point of view, see Ibrahim, \citeyear{Ibr1997}, and Celeux, Marin and
Robert, \citeyear{CalMarRob2006}).

\begin{prop} \label{prop:pik KL}
Consider the linear model $\cM$ specified in (\ref{eq:general
linear model}) with a \textup{NIGa}$(b,g(X^TX)^{-1}$, $d,a)$ prior on
$(\bet, \sdue)$ described in (\ref{eq:gNIGa M})--(\ref{eq:g-prior
M}) and a submodel $\cM_k$ specified in (\ref{eq:submodel mod
selection}). Conditionally on the assumption that $\sdue_k$ has the
same distribution as $\sdue$, that is,
\textup{IGa}$(d/2, a/2)$, the KL-prior on
$(\betk^*,\sduek)$ is given by
%
\begin{equation}\label{eq:g-prior KL}
\mbox{\textup{NIGa}}(\bet_k^*,\sdue_k; b^{\mathrm{KL}}_k,g(X_k^TX_k)^{-1},d,a ),
\end{equation}
with
%
\begin{equation}\label{eq:Ebatakl}
b_k^{\mathrm{KL}}=b_k+(X_k^T X_k)^{-1}(X_{k}^T X_{\setminus k})
b_{\setminus k},
\end{equation}
where $(b_k^T,b_{\setminus k}^T)$ is the
decomposition of $b^T=E(\beta)^T$ corresponding to $\cM_k$.
\end{prop}

\begin{pf}
Recalling that $\beta_k^{\perp}$ is a linear transformation of
$\bet$, it follows immediately that the distribution of
$\beta_k^{\perp}$ given $\sdue_k$ is normal. Now $
E(\beta_k^{\perp}|\sdue_k)=(X^T_k X_k)^{-1}X_k^TXE(\beta)=(X^T_k
X_k)^{-1}X_k^TX b$,\vspace*{-5pt}
and (\ref{eq:Ebatakl}) follows immediately rewriting $X=(X_k \vdots
X_{\setminus k})$ and\break $b^T=(b_k^T,b_{\setminus k}^T)$.
Furthermore,
$ \operatorname{Var}(\beta_k^{\perp}|\sdue_k)=g 
\sdue_k (X_k^T \* X_k)^{-1}W_k
$, where
$W_k=X_k^T P X_k(X_k^T X_k)^{-1}$
with $P=X(X^TX)^{-1}X^T$. Let now $M_{\setminus k}=(I-P_{\setminus
k})$, where
$P_{\setminus k}$ denote the projection matrix onto the column space of
$X_{\setminus k}$.
Using the
equality $P=I-M_{\setminus k}+M_{\setminus k}X_k(X^T_kM_{\setminus
k}X_k)^{-1}X_k^TM_{\setminus k}$ provided in Searle\break (\citeyear{Sea1982},
exercise 8, page 269), it
follows that $W_k=I$, which gives the result.
\end{pf}

Consider now the projection with respect to $(\beta_k^*,\sdue_k)$
whose corresponding expressions are provided in\break point~(iii) of
Lemma \ref{lemma}.
Notice that $\beta_k^{\perp}$ is unchanged relative to the
previous case; on the other hand $\sigma_k^{2 \perp} \geq\sdue$
since $Q_k(\bet)\geq0$. [This follows because $Q_k(\bet)$ can be
written as $w^Tw$ with $w=M_k X\bet$, using the fact that
$M_k$ is a projection matrix.] As a consequence the
KL-projection variance under $\cM_k$ will always exceed $\sdue$,
justifying the intuition that the variance under $\cM_k$ should
be larger to account for a greater lack of fit. This case
generalizes the simple linear regression example introduced shortly
before Section~\ref{sec:coherence nuisance}.

The KL-prior of $(\bet_k^*, \sdue_k)$, that is, that induced from
(\ref{eq:gNIGa M}) on ($\bet_k^{\perp},\sigma_k^{2 \perp}$), is
unfortunately not analytically available, because of the awkward
dependence of $\sigma_k^{2 \perp}$ on $(\bet,\sdue)$. Of course
one can easily simulate from the KL-prior on $(\bet_k^*, \sdue_k)$
using draws from the gNIGa prior on $(\bet, \sdue)$ and mapping
them into draws from $\pi_k^{\mathrm{KL}}$ through
$(\beta_k^{\perp}, \sigma_k^{2 \perp})$. However, we will not follow
this course of action and
derive an analytical approximation along the lines described in
Consonni, Guti\'errez-Pe\~na and
Veronese (\citeyear{ConGutVer2007}). Essentially, we
employ a conjugate
prior that minimizes the KL-divergence relative to the true $\pi
_k^{\mathrm{KL}}$. We
call the resulting prior the KL-conjugate approximation, but
for simplicity, we still identify it as $\pi_k^{\mathrm{KL}}$.
Specifically, we approximate the true KL-prior within the
conjugate gNIGa family, whose hyperparameters
$b^{\mathrm{KL}}_k,g^{\mathrm{KL}}_k,a^{\mathrm{KL}}_k,d^{\mathrm{KL}}_k$
are given in the following proposition.

\begin{prop}\label{prop:kl}
Consider the linear model $\cM$ specified in (\ref{eq:general
linear model}) with a \textup{NIGa}$(b,g(X^TX)^{-1}$, $d,a)$ prior on
$(\bet, \sdue)$ described in (\ref{eq:gNIGa M})--(\ref{eq:g-prior
M}) and a submodel $\cM_k$ specified in (\ref{eq:submodel mod
selection}). Then the KL-conjugate approximation prior on
$(\betk^*,\sduek)$ is the \textup{NIGa}$(\bkl\!,\gkl(X_k^TX_k)^{-1}$,
$\dkl,\akl)$
where the hyperparameters can be identified in the following way:
\begin{itemize}
\item If $\bnk=0$, they are the solutions of the
following system of equations:
%
\begin{eqnarray}
\bkl&=&b_k, \label{eq:bkl1}\\
\gkl&=&gE(R_k(\bet,\sdue)),\label{eq:gkl1}\\
\akl&=&\dkl\frac{a}{d} \frac{1}{E[R_k(\bet,\sdue)]},\label
{eq:akl1}\\
\label{eq:dkl1}
\psi(\dkl/2)-\log(\dkl/2)&=&\psi(d/2)-\log(d/2) \nonumber\\
&&{}+E\{\log[R_k(\bet
,\sdue)]\}\\
&&{}-\log\{E[R_k(\bet,\sdue)]\},\nonumber
\end{eqnarray}
where $R_k(\bet, \sdue)=(1+Q_k(\bet)/\sdue)^{-1}$,
and $\psi(\alpha)=\frac{\partial}{\partial\alpha}
\log(\Gamma(\alpha))$
is the digamma function.

\item If $\bnk\neq0$, they are \emph{approximately} the
solutions of the following system of equations:
%
\begin{eqnarray}
\bkl&=&b_k+(X_k^T X_k)^{-1}(X_{k}^T X_{\setminus k})
b_{\setminus k}, \label{eq:bkl2}\\
\gkl&=&\frac{g}{E[R_k(\bet,\sdue)^{-1}]}, \label{eq:gkl2}\\
\akl&=& \dkl\frac{a}{d}E[R_k(\bet,\sdue)^{-1}] \label{eq:akl2}
\end{eqnarray}
and
\begin{eqnarray}
\label{eq:dkl2}
&&\psi(\dkl/2)-\log(\dkl/2)\nonumber\\
&&\quad=\psi(d/2)-\log(d/2)\\
&&\qquad{} +
\frac{1}{2}\frac{\operatorname{Var}[R_k(\bet, \sdue
)^{-1}]}{E[R_k(\bet,
\sdue)^{-1}]^2}.\nonumber
\end{eqnarray}
\end{itemize}
The analytical expressions for $E[R_k(\bet,\sdue)]$,
$E[R_k(\bet,\break\sdue)^{-1}]$ and $\operatorname{Var}[R_k(\bet, \sdue)^{-1}]$
are given in Lemma \textup{\ref{lemmaA1}} in the \hyperref[app]{Appendix}.
\end{prop}

\begin{pf}
See the \hyperref[app]{Appendix}.
\end{pf}

Notice that both the expressions of $\bkl$ in Propositions \ref{prop:pik KL} and \ref{prop:kl} coincide
with that of $b_k^{\mathrm{UC}}$. Furthermore, (\ref{eq:bkl1})--(\ref{eq:dkl1}), as well as (\ref{eq:bkl2})--(\ref{eq:dkl2}),
do not admit a closed-form solution. Yet, a few results can be established
which we report without proof:
$\dkl<d$; $\dkl/d\rightarrow0$ for $d\rightarrow\infty$; $\akl
\rightarrow0$ for $d\rightarrow\infty$, whence $\akl<a$ for large $d$;
$E(\sigma^{-2}_k)=\dkl/\akl<d/a=\break E(\sigma^{-2})$, as expected.
Finally nested-coherence is satisfied on the space or regression
parameter, while it fails on the variance space. Moreover it can be established
empirically that nuisance-coherence fails.

\subsection{Information Paradox}
\label{subsec:info-par}
A major objection to the use of $g$-priors falls under the heading
of Information Paradox; see Liang et al. (\citeyear{LiaPauMolClyBer2008}) for a recent
discussion. Suppose
that the regression model $\cM_k$ is compared with the ``Null'' model
$\cM_0$ having no predictors. Assume the data
overwhelmingly support $\cM_k$, that is, $ \|\beta_k \|^2 = \beta_k^T
\beta_k \rightarrow
\infty$, so that the coefficient $R^2$ under $\cM_k$ tends to 1 and
$\mathit{SSR}_k=y^TM_ky \rightarrow0$.
Using a $g$-prior under both models with zero expectation for the
regression parameters
and $d_k=d$, $a_k=a$ and $g_k=g$, the Bayes
factor $B_{k0}$ of $\cM_k$ against $\cM_0$ remains bounded\break whereas
one would
expect it to diverge. However, the paradox does not necessarily arise
if we assume \emph{different}
$g$-priors under the two models as implied by the UC and KL-procedures,
as we now show.

First notice that
$ \beta_k^T\beta_k \rightarrow
\infty$ implies also $y^T y \rightarrow
\infty$. If $b=E(\bet)$ is independent of the data, for example; $b_0$
in (\ref{eq:m-default}), it can be easily checked using
(\ref{eq:Bks}) that $B_{k0}$ is asymptotic to
\begin{eqnarray}
\frac{ (y^TA y )^{(d_0+n)/2}}{ (1/(1+g_k)y^Ty )^{(d_k+n)/2}}\nonumber\\
\eqntext{\mbox{with } A= \biggl(I-\dfrac{g_0}{n(1+g_0)}J \biggr),}
\end{eqnarray}
where $I$ is the identity matrix and $J$ is the matrix with
all elements equal to 1.
Since $\lambda_{\min}\leq y^TAy/\break y^Ty \leq\lambda_{\max}$, where
$\lambda
_{\min}$ and $\lambda_{\max}$
are the smallest and largest eigenvalues of $A$, it follows that $y^TAy
= O(y^Ty)$ since
$\lambda_{\min}>0$.
As a consequence
the limiting behavior of $B_{k0}$ depends on the hyperparameters $d_0$
and $d_k$ deduced from
the specific compatible procedure. In the case of UC, we have
$g_k^{\mathrm{UC}}=g$ and $d_k^{\mathrm{UC}}=d+(p-p_k)<d+(p-1)=d_0^{\mathrm{UC}}$
and thus $B_{k0} \rightarrow\infty$ so that the paradox does not arise.
However, this result does not hold for the KL-procedure,
since $d_k^{\mathrm{KL}}>d_0^{\mathrm{KL}}$. The same conclusions can be obtained, using
similar arguments, if we assume $E(\bet)=\bar{b}=(\bar{y}, 0, \ldots
, 0)$.

Suppose now $E(\bet)=\hat{b}$, that is, the expectation of $\bet$ is fully
data-dependent.
In this case both $b_k^{\mathrm{UC}}$ and $b_k^{\mathrm{KL}}$
reduce to the OLS estimate of $\bet_k$ under $\cM_k$, that is,
$b_k^{\mathrm{UC}}=b_k^{\mathrm{KL}}=(X_k^TX_k)^{-1}X_k^Ty$, while
$b_0^{\mathrm{UC}}=b_0^{\mathrm{KL}}=\bar{y}$. Thus, from (\ref{eq:Bks}), $B_{k0}$ is
asymptotic to
%
\begin{eqnarray}\label{eq:Bk0}
C_{k0}\frac{ (a_0+y^TM_0
y )^{(d_0+n)/2}}{a_k^{(d_k+n)/2}}\nonumber\\[-8pt]\\[-8pt]
\eqntext{\mbox{with }
M_0= \biggl(I-\dfrac{1}{n}J \biggr).}
\end{eqnarray}
Under the UC procedure, only the hyperparameter $a_k^{\mathrm{UC}}$ can depend
on the data $y$ through
$\hat{b}$ [see (\ref{dkUC-1}) and (\ref{dkUC-2})], and we have
%
\begin{eqnarray}\label{eq:akUC-lim}
a_k^{\mathrm{UC}}&=&a+y^T M_k X_{\setminus k}^T (X_{\setminus k}^T M_k
X_{\setminus k})^{-1}X_{\setminus k}^T M_ky \nonumber\\
&=& a+y^T(P-P_k)y \\
&=&a+y^T(M_k-M)y \rightarrow a,\nonumber
\end{eqnarray}
recalling that
$b_{\setminus k}^{\mathrm{UC}}=\hat{\bet}_{\setminus k}=(X_{\setminus k}^T
M_k X_{\setminus k})^{-1}X_{\setminus k}^T M_ky$, and using
formula 3.98 on page 42 and Theorem A.45 on page 367 in Rao and
Toutemburg (\citeyear{RaoTou1999}). The result follows noting that
$y^T(M_k-M)y\rightarrow0$ because the $\mathit{SSR}$ of $\cM$
must be less than that of $\cM_k$ which tends to zero by hypothesis.
Thus $B_{k0}$ in (\ref{eq:Bk0}) trivially goes to infinity, since
$C_{k0} \rightarrow
\mbox{\textit{constant}}$ and $y^TM_0y \rightarrow\infty$, and there is no paradox.

Under the KL-procedure instead,
from (\ref{eq:dkl1}), (\ref{eq:gklapp}) and (\ref{eq:aklapp}), it
appears that the dependence of the\vspace*{2pt} hyperparameters
on the data happens only through $Q_k(\hat{\bet})$. Now
\begin{eqnarray*}
Q_k(\hat{\bet})&=&\frac{1}{n}\hat{\bet}^TX^TM_kX\hat{\bet}
=\frac
{1}{n}y^TPM_kPy\\
&=&\frac{1}{n}y^T(P-P_k)y=
\frac{1}{n}y^T(M_k-M)y
\end{eqnarray*}
which tends to zero as in (\ref{eq:akUC-lim}). Accordingly the
hyperparameters behave as constants
in the limit, and thus also in this case the information paradox does
not arise.

\section{Examples}\label{sec:examples}

In this section we present three examples in order to evaluate the
performance of the various priors discussed in Section
\ref{subsec:priors}. The first one considers the very simple situation
of testing
a normal model with a submodel
$\cM^*$ having mean zero: in this way different priors of
$\sigma^{*2}$ can be more easily compared. Features of the priors, and
their consequences on variable selection, are then assessed in a
more complex simulation study, and in a real data set
(Hald data), frequently analyzed in the literature.

\begin{figure*}

\includegraphics{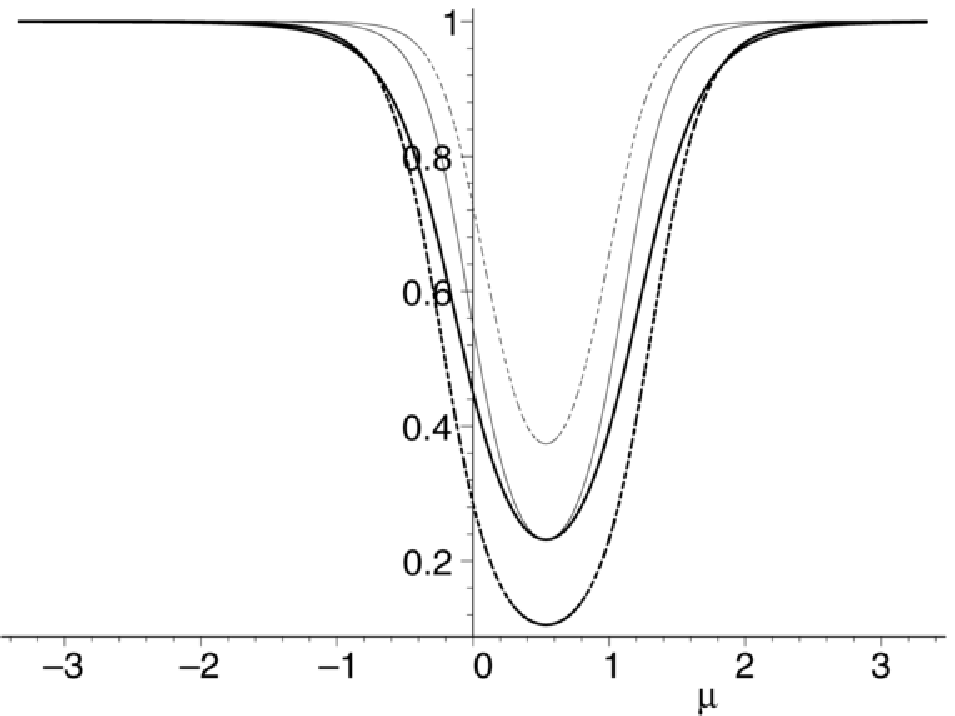}

\caption{Posterior probability of $\cM$ for hyperparameters $d=5$, $a=1$:
$p^{\mathrm{KL}}(\cM|y)$ dash thick, $p^{\mathrm{S}}(\cM|y)$ solid thin, $p^{\mathrm{UC}}(\cM|y)$
dash thin,
$p^{\mathrm{I}}(\cM|y)$ solid thick.}
\label{fig:pfinMa1d5}
\end{figure*}

\begin{figure*}[b]

\includegraphics{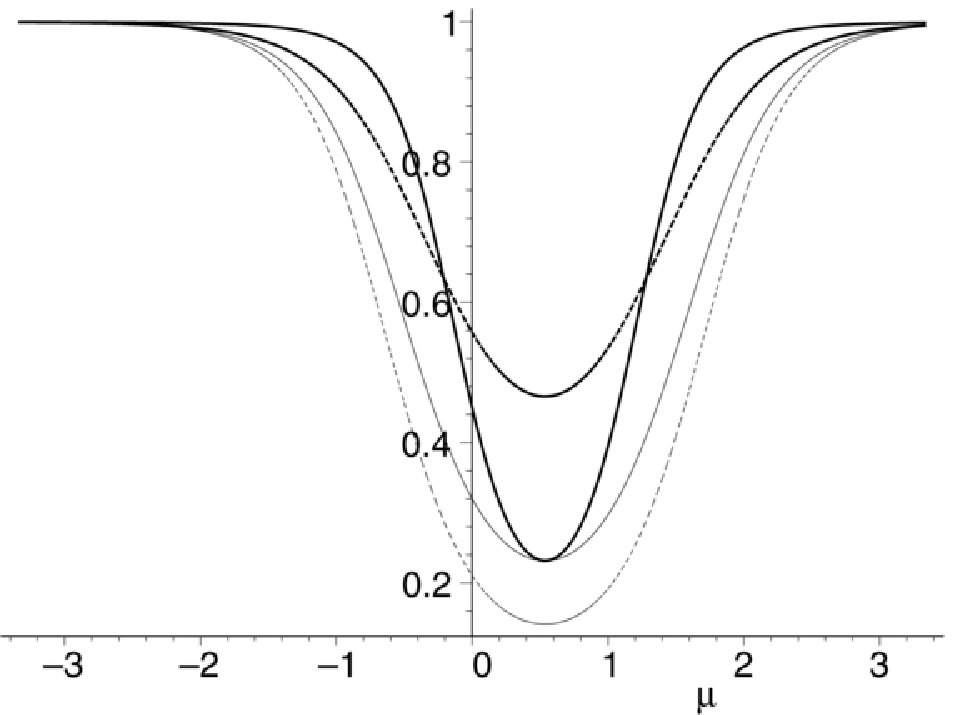}

\caption{Posterior probability of $\cM$ for hyperparameters
$d=3$, $a=25$: $p^{\mathrm{KL}}(\cM|y)$ dash thick, $p^{\mathrm{S}}(\cM|y)$ solid
thin, $p^{\mathrm{UC}}(\cM|y)$ dash thin, $p^{\mathrm{I}}(\cM|y)$ solid thick.}
\label{fig:pfinMa25d3}
\end{figure*}

\subsection{A Simple Illustration} \label{subsec:example1}

Consider the two models
\begin{eqnarray*}
\cM\dvtx \  y_i& =& \mu+\epsilon_i, \quad\epsilon_i
\stackrel{\mathrm{iid}}{\sim} \mbox{N}(0,\sigma^2),\\
\cM^*\dvtx \ y_i&=& \epsilon_i^*, \hspace*{8pt}\qquad\epsilon_i^*
\stackrel{\mathrm{iid}}{\sim}
\mbox{N}(0,\sigma^{*2}),
\end{eqnarray*}
with $i=1,\ldots,n$, and assume as a prior for $(\mu,
\sigma^2)$ under $\cM$ the following gNIGa:
$\pi(\mu|\sigma^2)=\mbox{N}(\mu; 0,g \sigma^2/\break n)$;
$\pi(\sigma^2)=\mbox{IGa}(\sigma^2; d/2,a/2)$.
If $St_n(\cdot; \eta, \Lambda, \nu)$ denotes an $n$-dimensional Student
t-distribution with expectation $\eta$, degrees of freedom $\nu$
and variance matrix $\nu\Lambda^{-1}/(\nu-2)$, $\nu>2$, the
marginal density of $y$ is
\begin{eqnarray*}
f(y)=St_n \biggl(y; 0,\frac{d}{a} \biggl(I-\frac{ g/n}{1+g}J \biggr),d \biggr).
\end{eqnarray*}

The submodel $\cM^*$ only requires a prior on
$\sigma^{*2}$. The Standard, UC and KL-procedures lead to priors $\ps$, $\puc$ and $\pkl$ for
$\sigma^{*2}$ which are all of type $\mbox{IGa}(d^*/2,a^*/2)$.
Specifically, one obtains
\begin{eqnarray}\label{eq:adS-adUC}
&(d^{\mathrm{S}}=d, \quad a^{\mathrm{S}}=a) ;& \nonumber\\[-8pt]\\[-8pt]
&(d^{\mathrm{UC}}=d+1, \quad a^{\mathrm{UC}}=a).&\nonumber
\end{eqnarray}
We consider also the typical improper
prior on $\sigma^2$ given by $\pi^{\mathrm{I}}(\sigma^2)\propto\sigma^{-2}$
which can be formally obtained from $\pi^{\mathrm{S}}$ setting $d=0$, $a=0$.
Consider now the KL-prior. A direct computation yields
$\sigma^{2 \perp}= \sigma^2+ \mu^2$,
which can also be deduced from (\ref{eq:KL-projection-sduek}) by
setting $P_k$
equal to the zero matrix since in this case $X_k$ is void, so that
$Q_k(\mu)=\mu^2$. The values of $d^{\mathrm{KL}}$ and $a^{\mathrm{KL}}$ can be
recovered from (\ref{eq:akl1}) and (\ref{eq:dkl1}).
For illustration, in the following we use three different values
of $(d,a)$, namely $(d=1, a=1)$, $(d=5, a=1)$, $(d=3, a=25)$
leading respectively to $(d^{\mathrm{KL}}=0.93, a^{\mathrm{KL}}=1.42)$,
$(d^{\mathrm{KL}}=3.38, a^{\mathrm{KL}}=1.03)$, $(d^{\mathrm{KL}}=2.36, a^{\mathrm{KL}}=29.98)$.

In order to appreciate the effect of the different priors, we
compute the posterior probability of the two models $\cM$ and
$\cM^*$. In particular assuming prior odds 1, we have
$\Pr(\cM|y)=1/(1+B^*)$, where $B^*=f^*(y)/f(y)$ is the Bayes
factor of $\cM^*$ versus $\cM$.
Notice that $f^*(y)=St_n (y; 0,(d^*/a^*)I, d^* )$ with $d^*$ and $a^*$
depending on the specific procedure.
We fix $n=g=25$ and perform a simulation study, generating
a vector $\epsilon$ from a multivariate standard normal distribution,
and set $y=\mu\iota_n+\epsilon$, where $\iota_n$ is the
$n$-dimensional unit vector. In
Figures \ref{fig:pfinMa1d5} and \ref{fig:pfinMa25d3} the posterior
probability of $\cM$ is plotted as a function of $\mu$.
Notice that the minimum of the curves does not occur at $\mu=0$,
because the generated errors in the simulation had a negative
mean of about $-0.5$.
Ideally the posterior probability curve should reach a minimum
close to zero for $\mu\approx0$ and then increase rapidly as
$\mu$ moves away from zero.
When $d=a$ all curves overlap to a large extent. Differences
emerge for unequal $a$ and $d$
with the curves corresponding to $\pi^{\mathrm{I}}$ and $\pi^{\mathrm{S}}$ occupying
intermediate positions, while those associated to $\pi^{\mathrm{KL}}$ and
$\pi^{\mathrm{UC}}$ represent ``extreme'' curves. A strong
sensitivity of
$\pi^{\mathrm{UC}}$ and $\pi^{\mathrm{KL}}$ is apparent and in particular when $a$ is
greater than $d$, $\pi^{\mathrm{UC}}$ favors $\cM^*$ most strongly, while
$\pi^{\mathrm{KL}}$ favors $\cM$ (and conversely when $d$ is greater than
$a$). For $a>d$, the curve
corresponding to $\pi^{\mathrm{S}}$ is somewhat flatter than that under $\pi^{\mathrm{I}}$.

We now consider the problem of model comparison from a predictive
viewpoint as described in\break  Gelfand and Ghosh (\citeyear{GelGho1998}); see also
Marriot, Spencer and
Pettitt (\citeyear{MarSpePet2001}). In the simple case corresponding to squared
error loss, each model $\cM_k$ is assigned a score $D^{(k)}$ made up
of two parts: an error sum of squares component $G^{(k)}$ and a
predictive variance component $P^{(k)}$,
%
\begin{eqnarray}\label{eq:criterioGG}
D^{(k)}&=&\frac{c}{c+1} G^{(k)}+P^{(k)}, \quad c>0,
\end{eqnarray}
where
\begin{eqnarray*}
G^{(k)}&=&\sum_{i=1}^n\bigl(\muik-y_i\bigr)^2, \\
P^{(k)}&=&\sum_{i=1}^n \sdueik;\\
\muik&=&E^{(k)}(\yirep|y), \\
\sdueik&=&\operatorname{Var}^{(k)}(\yirep|y).
\end{eqnarray*}

In the above setting $y^T=(y_1, \ldots,y_n)$ are the data, while
$\yirep$
represents a future replicate observation (the number of replicates
being equal to that of the data).
Model selection is achieved through a minimization of $D^{(k)}$ for
a given choice of $c$. The term $P^{(k)}$ represents a penalty which
aims at discouraging models that either strongly underfit or
overfit the data, because in both cases predictive variances will
tend to be inflated. Since our objective is to compare the
performances of the various priors under model $\cM^*$ we simply
need to evaluate $D^*$ for each distinct prior.

Consider first $\mu_i^*$. This is
\begin{eqnarray*}
\mu_i^*&=&E^*(\yirep|y)=E^*[E^*(\yirep|y,\sigma
^{2*})|y]\\
&=&E^*[E^*(\yirep|\sigma^{2*})|y]=0,
\end{eqnarray*}
since under $\cM^*$ each observation has expectation zero,
conditionally on $\sigma^{2*}$.
As a consequence $D^{*}=P^{*}+\sum_{i=1}^n y^2_i$, and thus only the
term $P^*$ matters for comparison purposes. Now
\begin{eqnarray*}
\sigma^{2*}_i&=&\operatorname{Var}^*(\yirep|y)=E^*[\operatorname
{Var}^*(\yirep|y,\sigma
^{2*})|y]\\
&=&E^*(\sigma^{2*}|y)=
\frac{a_n^*}{d_n^*-2}, \quad d_n^*-2>0,
\end{eqnarray*}
since under each prior
the posterior distribution of $\sigma^{*2}_i$ is IGa$(d_n^*/2,
a_n^*/2)$, with $d_n^*=d^*+n$, and $a_n^*=a^*+\sum_{i=1}^n y^2_i$.
In conclusion the predictive criterion of Gelfand and Ghosh
(\citeyear{GelGho1998}) suggests to base model comparison on
$P^*=n a_n^*/(d_n^*-2)$.

From (\ref{eq:adS-adUC}), it is immediate to conclude that
$P^{\mathrm{UC}}<P^{\mathrm{S}}$ so that
$\puc$ supports $\cM^*$ more than $\ps$.
On the other hand, since $d^{\mathrm{KL}}<d$ it follows that
$P^{\mathrm{KL}}>P^{\mathrm{S}}$ whenever $a^{\mathrm{KL}}>a^{\mathrm{S}}$ (calculations show that this
occurs for moderate values of $d$, specifically $d<5.45$); in other
words the KL-prior would tend to favor $\cM^*$ less
than $\ps$. These conclusions are broadly in accord with
the curves describing $P(\cM|y)$ depicted in Figures~\ref{fig:pfinMa1d5} and \ref{fig:pfinMa25d3}.

\subsection{Simulation Study}

As a second example, we consider a simulation study along the lines presented
in George and McCulloch (\citeyear{GeoMcC1993}), Raftery, Madigan and
Hoeting (\citeyear{RafMadHoe1997}) and Fern\'{a}ndez, Ley and Steel (\citeyear{FerLeySte2001}).
We consider $p=6$ predictors, the constant plus $(X_1,\ldots,X_5)$ and
$n=30$ observations. Let $Z_j$, $j=1,\ldots,5$ be independent
$n$-dimensional vectors,
whose components are independent standard normal variables, and
set
\begin{eqnarray*}
X_1&=&Z_1,\quad X_2=Z_2,\quad X_3=Z_3, \\
(X_4,X_5)&=&(X_1,X_2)(0.3\ 0.7)^T(1\ 1)+(Z_4,Z_5).
\end{eqnarray*}
In this way there is a correlation
between the first two predictors and the last two.
We generate the response $y$ according to three different
models:
%
\begin{eqnarray}
\qquad\cM_{1}\dvtx y&=&C+2.5 \epsilon,\\
\qquad\cM_{2}\dvtx y&=&C+2X_1-X_3+1.5X_5+2.5\epsilon,\\
\qquad\cM_{3}\dvtx y&=&C+2X_1-X_3+X_4+1.5X_5+2.5\epsilon,
\end{eqnarray}
where $C$ is a fixed constant and the $n$
elements of $\epsilon$ are independent standard normal variables.
In particular, the case in which the data were generated from
$\cM_1$ was analyzed in a frequentist way by Freedman (\citeyear{Fre1983}). He
showed that, under this ``null model,'' standard
variable selection procedures, such as stepwise regression, may
lead to misleading results, for example, retaining a subset of predictors
with a highly significant $F$-statistic and reasonably high $R^2$.

\begin{table*}
\caption{Frequency of correct identification of the true model $\cM_{i}$ $(i=1,2,3)$
with $g=n=30$ for various
compatible priors and different choices of $(d,a)$ and $E(\beta)$}
\label{tab:Sim-M01}
\begin{tabular*}{\textwidth}{@{\extracolsep{\fill}}lccccccccccccc@{}}
\hline
& & \multicolumn{3}{c}{$\bolds{\pi^{\mathbf{KL}}}$}&
\multicolumn{3}{c}{$\bolds{\pi^{\mathbf{S}}}$} &
\multicolumn{3}{c}{$\bolds{\pi^{\mathbf{UC}}}$} &
\multicolumn{3}{c@{}}{$\bolds{\pi^{\mathbf{I}}}$} \\[-5pt]
&&\multicolumn{3}{c}{\hrulefill}&\multicolumn{3}{c}{\hrulefill}&\multicolumn{3}{c}{\hrulefill}&\multicolumn{3}{c@{}}{\hrulefill}\\
$\bolds{d}$ & $\bolds{a}$ & $\bolds{b_0}$ & $\bolds{{\bar b}}$ & $\bolds{\hat{b}}$ & $\bolds{b_0}$ & $\bolds{{\bar b}}$ &
$\bolds{\hat{b}}$
& $\bolds{b_0}$ & $\bolds{{\bar b}}$ & $\bolds{\hat{b}}$ & $\bolds{b_0}$ & $\bolds{{\bar b}}$ & $\bolds{\hat{b}}$\\
\hline
$\cM_1$ true model \\
\phantom{0}0 & \phantom{0}0 & & & & & & & & & &0.60 &0.56 &0.54 \\
\phantom{0}1 & \phantom{0}1 & 0.24 & 0.40 & 0.24 & 0.56 & 0.54 & 0.52 & 0\phantom{.00} & 0\phantom{.00} & 0.76 & & & \\
\phantom{0}1 & 10 & 0.08 & 0.24 & 0.48 & 0.64 & 0.56 & 0.56 & 0.32 & 0.26 & 0.64 & & & \\
\phantom{0}5 & \phantom{0}5 & 0.26 & 0.44 & 0.24 & 0.50 & 0.50 & 0.48 & 0.06 & 0.06 & 0.86 & & & \\
10 & \phantom{0}1 & 0.34 & 0.48 & 0.30 & 0.40 & 0.36 & 0.36 & 0\phantom{.00} & 0\phantom{.00} & 0.96 & & & \\
10 & 50 & 0.04 & 0.06 & 0\phantom{.00} & 0.56 & 0.54 & 0.52 & 0.46 & 0.42 & 0.60 & & & \\[6pt]
$\cM_2$ true model & & & & & & & & & & & & &\\
\phantom{0}0 & \phantom{0}0 & & & & & & & & & &0.46 &0.56&0.60 \\
\phantom{0}1 & \phantom{0}1 & 0.70 & 0.60 & 0.66 & 0.48 & 0.58 & 0.60 & 0.68 & 0.68 & 0.32 & & & \\
\phantom{0}1 & 10 & 0.70 & 0.66 & 0.64 & 0.42 & 0.54 & 0.58 & 0.58 & 0.62 & 0.56 & & & \\
\phantom{0}5 & \phantom{0}5 & 0.68 & 0.60 & 0.68 & 0.58 & 0.60 & 0.62 & 0.64 & 0.64 & 0.18 & & & \\
10 & \phantom{0}1 & 0.66 & 0.52 & 0.60 & 0.62 & 0.64 & 0.66 & 0\phantom{.00} & 0\phantom{.00} & 0.02 & & & \\
10 & 50 & 0.66 & 0.68 & 0.68 & 0.50 & 0.58 & 0.60 & 0.58 & 0.62 & 0.58 & & & \\[6pt]
$\cM_3$ true model & & & & & & & & & & & & &\\
\phantom{0}0 & \phantom{0}0 & & & & & & & & & &0.26 &0.38 &0.54 \\
\phantom{0}1 & \phantom{0}1 & 0.64 & 0.44 & 0.66 & 0.26 & 0.42 & 0.54 & 0.68 & 0.68 & 0.26 & & & \\
\phantom{0}1 & 10 & 0.74 & 0.54 & 0.52 & 0.24 & 0.36 & 0.05 & 0.30 & 0.42 & 0.54 & & & \\
\phantom{0}5 & \phantom{0}5 & 0.54 & 0.32 & 0.50 & 0.40 & 0.54 & 0.56 & 0.64 & 0.64 & 0.04 & & & \\
10 & \phantom{0}1 & 0.22 & 0.16 & 0.66 & 0.56 & 0.56 & 0.60 & 0\phantom{.00} & 0\phantom{.00} & 0\phantom{.00} & & & \\
10 & 50 & 0.74 & 0.56 & 0.60 & 0.34 & 0.48 & 0.54 & 0.50 & 0.54 & 0.52 & & & \\
\hline
\end{tabular*}
\end{table*}

In order to compare the different priors, we consider the Bayes factor
for each submodel versus the full model with six predictors (including
the constant)
for 50 simulated data sets
and report the frequency of times in which the highest Bayes
factor is associated to the correct model (i.e., the model which
has generated the data).
We fix $g=n$ and for each choice of
$E(\beta)$, namely $b_0, \bar{b}, \hat{b}$ [see (\ref{eq:m-default})]
check the robustness of the various
priors to the choice of the hyperparameters
$(d,a)$ of the inverse-gamma distribution on $\sdue$ (each time
leaving unchanged the values of the predictors).

We can summarize our results, which are in part reported in Table
\ref{tab:Sim-M01}, as
follows:
\begin{longlist}[(iii)]
\item[(i)]
$\pi^{\mathrm{UC}}$ appears to be the least robust prior relative to the
various choices of $E(\beta)$ and $(d,a)$;
this is consistent with the fact that the marginal of the data under
$\pi^{\mathrm{UC}}$ is
more peaked on its expectation; see the discussion in
Section \ref{subsec:example1}.
Its frequency of correct model identification can reach very low
values especially when $d$ exceeds $a$, in accord with the
fact that as $d$ increases relative to $a$ larger models
receive greater support under $\pi^{\mathrm{UC}}$; see Figure
\ref{fig:pfinMa1d5}.
To provide an explanation of this phenomenon, consider the Bayes
factor $B_k$ of the submodel $\cMk$ versus the full model $\cM$. If
the prior under $\cMk$ is obtained through UC, then calculations
show that
%
\begin{eqnarray}\label{eq:savage_ratio}
B_k=\frac{\pi(\bet_{\backslash k}=0 |y)}{\pi(\bet_{\backslash k}=0)},
\end{eqnarray}
where $\pi(\bet_{\backslash k}=0 |y)$ and $\pi(\bet_{\backslash
k}=0)$ are respectively the marginal posterior and prior density
of $\bet_{\backslash k}$, evaluated at the value $0$. The expression
(\ref{eq:savage_ratio}) for $B_k$ is known as ``Savage's
density ratio''; see, for example, O'Hagan and Forster (\citeyear{OHaFor2004},
Section~7.16). Now if the data are at least moderately more informative than
the prior, the numerator will be essentially dominated by the
likelihood, and thus will be fairly robust to prior specifications,
while this does not clearly occur for the denominator. In
particular, if $d$ increases relative to $a$, the distribution of
$\sdue$ tends to concentrate on smaller values, so that the marginal
of $\bet_{\backslash k}$ becomes more peaked around the mode (which
coincides with 0 under $b_0$ or $\bar{b}$), thus lowering $B_k$, and
supporting $\cM$ more than $\cMk$.

\item[(ii)]
$\pi^{\mathrm{KL}}$ is reasonably robust and shows good performance, save
when the generating model corresponds to the ``null model'' $\cM_1$ and $a$ is large (this is in accord with the fact
exhibited in Figure \ref{fig:pfinMa25d3} that for large $a$ bigger
models are preferred under $\pi^{\mathrm{KL}}$).
\item[(iii)] $\pi^{\mathrm{S}}$ and $\pi^{\mathrm{I}}$ exhibit a relatively similar
behavior, as already remarked in the previous section, and have a
better performance than the other priors at identifying the ``null
model.''
\end{longlist}

Overall, the frequency of correct model identification is
comparable, or even superior, to similar investigations carried
out in a Bayesian
framework, although using different model
choice criteria and different priors; see Marriot, Spencer and
Pettitt (\citeyear{MarSpePet2001}).

\begin{table*}[b]
\caption{Posterior probability of top four models with $g=n=13$
($\gamma=0.07$) and $g=9$ ($\gamma=0.1$)
for various compatible priors and different choices of $E(\beta)$; in
first column is Ibrahim's results}
\label{tab:Hald-top-model}
\begin{tabular*}{\textwidth}{@{\extracolsep{\fill}}lccccccccc@{}}
\hline
&&\multicolumn{2}{c}{$\bolds{\pi^{\mathbf{KL}}}$} &
\multicolumn{2}{c}{$\bolds{\pi^{\mathbf{S}}}$} &
\multicolumn{2}{c}{$\bolds{\pi^{\mathbf{UC}}}$} &
\multicolumn{2}{c@{}}{$\bolds{\pi^{\mathbf{I}}}$} \\[-5pt]
& &\multicolumn{2}{c}{\hrulefill}&\multicolumn{2}{c}{\hrulefill}&\multicolumn{2}{c}{\hrulefill}&\multicolumn{2}{c@{}}{\hrulefill}\\
\textbf{Model} & $\bolds{\pi^{\mathbf{Ibr}}}$& $\bolds{{\bar b}}$ & $\bolds{\tilde{b}}$ & $\bolds{{\bar b}}$ & $\bolds{\tilde{b}}$
& $\bolds{{\bar b}}$ & $\bolds{\tilde{b}}$ & $\bolds{{\bar b}}$ & $\bolds{\tilde{b}}$\\
\hline
$g=13$ \\
$\{1,2\}$ & &0.175 &0.203 &0.340 &0.290 & 0.276&0.293 &0.329 &0.271 \\
$\{1,4\}$ & & & & & & & &0.221 & \\
$ \{1,2,3\}$ & &0.181 &0.227 &0.145 &0.207 & 0.167&0.211 &0.112 &0.213 \\
$\{1,2,4\}$ & &0.184 &0.234 &0.151 &0.220 & 0.174&0.223 &0.114 &0.229 \\
$ \{1,3,4\}$ & &0.169 &0.174 &0.127 &0.155 & 0.147&0.146 & &0.153 \\[3pt]
Total & & 0.709 &0.838 & 0.763&0.872 & 0.764&0.873 &0.776 &0.866 \\[6pt]
$g=9$ & & & & & & & & & \\
$\{1,2\}$ & 0.272 &0.217 &0.210 &0.310 &0.262 &0.238 &0.268 & 0.294&0.248 \\
$\{1,4\}$ & &0.171 & &0.165 & & & & 0.219& \\
$\{1,2, 3\}$ & 0.215 &0.157 &0.230 &0.143 &0.215 &0.171 &0.222 & 0.111&0.219 \\
$\{1,2, 4\}$ & 0.214 &0.156 &0.216 &0.143 &0.209 &0.171 &0.217 & 0.111&0.213 \\
$\{1,3,4\}$ & 0.164 & &0.173 & &0.163 &0.153 &0.157 & &0.159 \\[3pt]
Total & 0.865 & 0.701 &0.829 &0.761 &0.852 &0.733 &0.864 & 0.735&0.839 \\
\hline
\end{tabular*}
\end{table*}

\subsection{Hald Data}\label{subsec:Hald}

Our third example involves the Hald data, often analyzed in the
literature, in order to evaluate model selection procedures; see, for
instance, Draper and Smith (\citeyear{DraSmi1981}). It consists of 13 observations on
one response variable with four predictors.
A specific feature of this
data set is represented by the strong correlation between $X_1$ and
$X_3$ and between $X_2$ and $X_4$. We consider all the possible 16
models in which the constant term is always included.

A detailed subjective Bayesian analysis of this data set has been performed
in Laud and Ibrahim (\citeyear{LauIbr1995}, \citeyear{LauIbr1996}) and Ibrahim (\citeyear{Ibr1997}), especially
in terms of prior specification.
We follow Laud and Ibrahim (\citeyear{LauIbr1995}) and fix a prior on $(\beta, \sdue)$
under the full model which is a NIGa$(\tilde{b},g(X^TX)^{-1},25,125)$ with
$E(\bet)=\tilde{b}=(X^TX)^{-1}X^T\eta$, where $\eta$ is a
subjective prediction for $y$ given by
$\eta=(79,77,104,90,99,108,105,73,93,\break111,88,115,113).$
We also report the value
$\gamma=1/(g+1)$, which represents a weight on the
prior guess~$\eta$. Notice that the choice of $d=25$ and $a=125$
implies $E(\sigma^{-2})=0.2$ and $\Pr(\sigma^{-2}<0.5)\approx
0.95$.

Table \ref{tab:Hald-top-model} summarizes the results of a Bayesian
analysis using the conventional
value $g=n=13$, as well as $g=9$ (Ibrahim's choice) which correspond
to\break
weights $\gamma=0.07$, respectively 0.10, representing weak prior information.
Moreover we consider
two choices for $E(\beta)$, namely $\bar{b}$ and $\tilde{b}$. We do
not report explicitly results for $E(\beta)=b_0$ because
posterior model probabilities are relatively more diffuse and no subset of
models emerges as a clear winner.
The column $\pi^{\mathrm{Ibr}}$ reports the results obtained in
Ibrahim (\citeyear{Ibr1997}) which assumes a fixed
$\sigma^{-2}=0.2$.
The highest probability is given to model $\{1,2\}$ under all
priors, save for $\pi^{\mathrm{KL}}$ that indicates a slight preference for
more complex
models, for example, $\{1,2,4\}$ for $g=13$. Overall there is broad agreement
with standard frequentist
model selection procedures as reported in Laud and Ibrahim (\citeyear{LauIbr1995},
Table 1).

We also performed a sensitivity analysis (not reported here) with
respect to $\gamma$ $(0.01 \leq\gamma\leq0.95)$ for the two choices
$E(\beta)=b_0$, respectively
$\tilde{b}$, in order to make a comparison with the results of Tables
2 and 3 of Ibrahim (\citeyear{Ibr1997}).
The results are appreciably sensitive to the choice of $b_0$ or
$\tilde{b}$, although this fact is definitely less manifest for
the prior $\pi^{\mathrm{Ibr}}$ (under which, however, $\sdue$ is assumed
fixed). Overall it is confirmed that the choice of $b_0$ is the least
satisfactory, as it tends to shift posterior model probability
toward ``extreme'' models, such as the null or full\vspace*{2pt}
model, when $\gamma$ approaches either boundary. On the other hand,
under $\tilde{b}$ the results are fairly
insensitive to the choice of $\gamma$ as far as the identification of
the top model is concerned, which is usually $\{1,2\}$, and either
$\{1,2,3\}$ or $\{1,2,4\}$. In particular $\pi^{\mathrm{KL}}$ exhibits a
high stability, with respect to $\gamma$, of the posterior probability
mass on the top model which always contains three predictors.

The Hald data have been also analyzed in a Bayesian objective
framework, in particular by Berger and Pericchi (\citeyear{BerPer1996b}) using
intrinsic Bayes factor, and by Casella and Moreno (\citeyear{CasMor2006}) and Moreno
and Giron (\citeyear{MorGir2007}) using intrinsic priors.
The models they identify are essentially those exhibited as most probable
in Table \ref{tab:Hald-top-model}. However, under their approach,
model $\{1,2\}$ receives a posterior
probability in excess of 50\%.
Based on an objective predictive approach, Barbieri and Berger (\citeyear{BarBer2004})
develop a theory for model choice.
They show that the optimal model is not necessarily the highest
posterior probability model, but rather the
``median probability model.'' For the Hald data the latter
is represented by $\{1,2,4\}$ which,
curiously, is also the model with the highest posterior
probability under the KL-prior with $g=n$; see Table~\ref{tab:Hald-top-model}.

\section{Discussion} \label{sec:discussion}

For a given proper prior on the parameter space of a full model, we
reviewed and analyzed procedures for the
specification of prior distributions on the parameter space of a
collection of submodels. We presented two interpretations of nested
models, in order to
explicate more naturally the rationale of each procedure.
In particular, we investigated four methods
for the specification of a compatible prior under a
submodel, namely marginalization, usual and Jeffreys conditioning
and Kullback--Leibler projection.
Next, each procedure was evaluated
from two perspectives, nuisance- and nested-coherence.
Given a full linear model with a normal inverted gamma $g$-prior on the
parameters,
we considered the problem of variable selection,
and applied the above procedures for the construction of priors under
each submodel $\cM_k$.
For completeness we also considered, for each $\cM_k$, a $g$-prior on
the regression
parameters combined with an inverted gamma $(d,a)$ distribution on
$\sigma_k^2$,
labeled $\pi^{\mathrm{S}}$, as well as a conventional
improper prior on $\sigma_k^2$, identified with $\pi^{\mathrm{I}}$.

Three examples were used to illustrate the behavior of the various
procedures for prior
specification, leading to the conclusions
that results
are quite sensitive to the choice of the hyperparameters.
Overall the improper prior $\pi^{\mathrm{I}}$ performs comparably to the standard
prior $\pi^{\mathrm{S}}$, when $d$ and $a$ are similar.
The usual conditioning prior $\pi^{\mathrm{UC}}$, despite its theoretically
attractive coherence properties exhibited in Propositions \ref{prop:nuisance coherence} and \ref{prop:nested coherence},
shows remarkable sensitivity to the choice of the hyperparameters,
oscillating between highly simple and complicated models.
The Kullback--Leibler projection prior exhibits a performance which is
comparable or superior to that of $\pi^{\mathrm{S}}$ when using the OLS estimate as
prior expectation on $\beta$, provided that the true model is
not very close to the ``null'' model with no
predictors. This is consistent with the general attitude of the
KL-prior to favor more complex models.

When the goal of model choice is prediction, one might consider
orthogonalizing the matrix of predictors, as in Clyde, DeSimone and
Parmigiani (\citeyear{ClyDeSPar1996}). In this case a $g$-prior on the regression
coefficient under the full model admits a diagonal variance
matrix. As a consequence the M, UC and KL-procedures would generate the
same prior under each submodel $\cM_k$ conditionally on
$\sigma^2_k$; yet they would imply distinct priors for the
variance. We remark, however, that this approach cannot be
implemented in a variable selection problem, where the focus is
on the original predictors.

Consistency of the posterior distribution on model space under
different choices of the hyperparameter $g_k^*$ in the gNIGa
prior (\ref{eq:gNIGa Mk}), with $d_k=d$ and $a_k=a$, has been
recently discussed in Fern\'{a}ndez, Ley and Steel (\citeyear
{FerLeySte2001}). They
prove, under mild conditions, that consistency obtains under both
the standard and improper priors $\ps$ and $\pi^{\mathrm{I}}$. Using similar
arguments one can prove that the same result holds for the UC
procedure under $b_0$ and $\bar{b}$, defined in (\ref{eq:m-default}).
As far as $\pi^{\mathrm{KL}}$ is concerned the limiting probability of
model $\cM_k$ is zero provided the true model is not nested within
$\cM_k$; on the other hand when $\cM_k$ is moderately larger than
the true model this result may fail, and $\pi^{\mathrm{KL}}$ may lead to
choose slightly overparametrized models.

It is well known that a standard use of $g$-priors for variable
selection cannot be recommended because it suffers from the
information paradox. However, our analysis shows that, when
$g$-priors under submodels are \textit{derived} using compatibility
criteria, the
paradox either does not arise (UC procedure), or can be avoided
(KL-procedure) through a suitable choice of the initial
hyperparameters.

Recent contributions in the area of linear models (see Liang
et al., \citeyear{LiaPauMolClyBer2008} and Bayarri and Garcia-Donato, \citeyear{BayGar2007}), suggest
to use a noninformative improper prior on the nuisance parameter and
a proper mixture of \mbox{$g$-priors} on the regression coefficients.
It would be interesting to apply the methods discussed in this paper to
the latter distribution of the regression coefficients in order to
derive a compatible mixture of
$g$-priors under the various submodels.

\begin{appendix}

\section*{Appendix}\label{app}
\renewcommand{\theequation}{\arabic{equation}}

\begin{pf*}{Proof of Proposition \ref{prop:nuisance coherence}}
Assume that the sampling distribution under model $\cM$ is
$\{ f(y|\lambda,\delta,\phi) \}$, where $\delta$ is the nuisance
parameter. Then, for a given prior $\pi(\lambda,\delta,\phi)$,
the integrated model $\cIM$ has sampling
distribution $f(y|\lambda,\phi)=\int
f(y|\lambda,\delta,\phi)\pi(\delta|\lambda,\break\phi)\,d\delta$,
while the corresponding integrated submodel $\cIM^*$ has density
$ f^*(y|\lambda)=f(y|\lambda,\phi=\phi_0)$.
Let the prior under $\cIM$ be
$\pi_{\cIM}(\lambda,\phi)=\pi(\lambda,\phi)$, that is, the
marginal distribution of $(\lambda,\phi)$ under $\pi$.
Consider now a procedure to construct a prior under a submodel.
Let
%
\begin{eqnarray}\label{eq:f*cM*}
&&f^*_{\cM^*}(y)\nonumber\\
&&\quad=\int\int
f^*(y|\lambda,\delta)\pi^*_{\cM^*}(\lambda,\delta)\,d\lambda
\,d\delta
\\
&&\quad= \int\int
f(y|\lambda,\delta,\phi=\phi_0)\pi^*_{\cM^*}(\lambda,\delta
)\,d\lambda\,d\delta\nonumber
\end{eqnarray}
and
%
\begin{eqnarray}\label{eq:f*cIM*}
&&f^*_{\cIM^*}(y)\nonumber\\
&&\quad=\int f^*(y|\lambda)\pi^*_{\cIM^*}(\lambda)\,d\lambda
\nonumber\\[-8pt]\\[-8pt]
&&\quad= \int\biggl\{ \int f(y|\lambda,\delta,\phi=\phi_0)
\pi(\delta|\lambda,\phi=\phi_0) \,d\delta\biggr\}\nonumber\\
&&\phantom{\int}\qquad{}\cdot\pi^*_{\cIM^*}(\lambda)\,d\lambda,\nonumber
\end{eqnarray}
where
$\pi^*_{\cM^*}(\lambda,\delta)$ is the output of the
procedure applied to $(\cM,\cM^*)$ starting from
$\pi(\lambda,\delta,\phi)$, while $\pi^*_{\cIM^*}(\lambda)$
is the output of the procedure applied to $(\cIM,\cIM^*)$ starting
from $\pi(\lambda,\phi)$.

(i) Recall that
$\pi^{\mathrm{UC}}_{\cM^*}(\lambda,\delta)=\pi(\lambda,\delta|\phi=\phi_0)$
and consider $\pi^{\mathrm{UC}}_{\cIM^*}$. We have
$\pi^{\mathrm{UC}}_{\cIM^*}(\lambda)=\pi_{\cIM}(\lambda|\phi=\phi_0)
=\pi(\lambda|\phi=\phi_0)$.
As a consequence we get from (\ref{eq:f*cM*})
%
\begin{eqnarray}\label{eq:UCf*cM*}
&&\quad f^{\mathrm{UC}}_{\cM^*}(y)\nonumber\\
&&\qquad= \int\biggl\{ \int
f(y|\lambda,\delta,\phi=\phi_0)\pi(\delta|\lambda,\phi=\phi
_0)\,d\delta\biggr\}\\
&&\phantom{\int}\quad\qquad{}\cdot \pi(\lambda|\phi=\phi_0)\,d\lambda,\nonumber
\end{eqnarray}
while from (\ref{eq:f*cIM*}) we get
%
\begin{eqnarray}\label{eq:UCf*cIM*}
&&\quad f^{\mathrm{UC}}_{\cIM^*}(y)\nonumber\\
&&\qquad=\int
\biggl\{ \int f(y|\lambda,\delta,\phi=\phi_0)
\pi(\delta|\lambda,\phi=\phi_0) \,d\delta\biggr\}\\
&&\phantom{\int}\quad\qquad{}\cdot
\pi(\lambda|\phi=\phi_0)\,d\lambda,\nonumber
\end{eqnarray}
and the two densities clearly coincide.

(ii) Recall that
$\pi^{\mathrm{M}}_{\cM^*}(\lambda,\delta)=\pi(\lambda,\delta)$. Consider
now
$\pi^{\mathrm{M}}_{\cIM^*}(\lambda)$: this is the marginal of
$\pi_{\cIM}(\lambda,\delta)$; the latter, however, coincides with the
marginal $\pi(\lambda,\delta)$
under the prior $\pi(\lambda,\delta,\phi)$ by definition of integrated
model. We therefore obtain
$\pi^{\mathrm{M}}_{\cIM^*}(\lambda)=\pi_{\cIM}(\lambda)=\pi(\lambda)$.
From (\ref{eq:f*cM*}) we get
\begin{eqnarray*}
f^{\mathrm{M}}_{\cM^*}(y)&=&\int\int
f(y|\lambda,\delta,\phi=\phi_0)\pi(\delta|\lambda)\pi(\lambda)\,d\delta\,d\lambda,
\end{eqnarray*}
while from (\ref{eq:f*cIM*}) we get
\begin{eqnarray*}
&&f^{\mathrm{M}}_{\cIM^*}(y)\\
&&\quad=\int\biggl\{ \int f(y|\lambda,\delta,\phi=\phi_0)
\pi(\delta|\lambda,\phi=\phi_0) \,d\delta\biggr\}\\
&&\qquad\phantom{\int}
{}\cdot\pi(\lambda)\,d\lambda.
\end{eqnarray*}
Inspection of $f^{\mathrm{M}}_{\cM^*}(y)$ and
$f^{\mathrm{M}}_{\cIM^*}(y)$ reveals that if $\delta$ is conditionally
independent of $\phi$ given $\lambda$, the two densities are equal.

(iii) Recall that
\begin{eqnarray*}
\pi^{\mathrm{JC}}_{\cM^*}(\lambda,\delta)\propto
\pi(\lambda,\delta|\phi=\phi_0)
\frac{j_{\cM^*}(\lambda,\delta)}{j_{\cM}(\lambda,\delta,\phi_0)},
\end{eqnarray*}
where the $j$-functions are the Jeffreys priors. Passing to
the integrated model we therefore obtain
\begin{eqnarray*}
\pi^{\mathrm{JC}}_{\cIM^*}(\lambda)\propto\pi_{\cIM}(\lambda|\phi=\phi_0)
\frac{j_{\cIM^*}(\lambda)}{j_{\cIM}(\lambda,\phi_0)}.
\end{eqnarray*}
Let
\begin{eqnarray*}
h(\lambda,\delta)=\frac{j_{\cM^*}(\lambda,\delta)}{j_{\cM
}(\lambda
,\delta,\phi_0)},
\quad
g(\lambda)=\frac{j_{\cIM^*}(\lambda)}{j_{\cIM}(\lambda,\phi_0)}.
\end{eqnarray*}
Clearly, if $h(\lambda,\delta)\propto g(\lambda)$, then
$f^{\mathrm{JC}}_{\cM^*}(y)$ and $f^{\mathrm{JC}}_{\cIM^*}(y)$ have a
representation as in (\ref{eq:UCf*cM*}), respectively
(\ref{eq:UCf*cIM*}), with the integrand in each case multiplied
by $g(\lambda)$, and therefore they must coincide.
\end{pf*}

\begin{pf*}{Proof of Proposition \ref{prop:nested coherence}}
Start with the M procedure. Notice that $\pi^{**}_{\cM}(\lambda)=\pi
(\lambda)$. On the other hand
$\pi^{**}_{\cM^*}(\lambda)=\pi^{*}(\lambda)$, where $\pi
^*(\lambda)$ is
the marginal prior
on $\lambda$ under $\pi^*(\lambda,\phi_2)$; but the latter is under M
equal to
$\pi(\lambda,\phi_2)$, whence $\pi^{**}_{\cM^*}(\lambda)=\pi
(\lambda)$,
thus establishing the result.

Consider now the UC procedure. We have
$\pi^{**}_{\cM}(\lambda)=\pi(\lambda|\phi_1=\phi_1^0,\phi
_2=\phi_2^0)$.
On the other hand
\begin{eqnarray*}
\pi^{**}_{\cM^*}(\lambda)&=&\pi^*(\lambda|\phi_2=\phi_2^0)
=\frac{\pi^*(\lambda,\phi_2=\phi_2^0)}{\pi^*(\phi_2=\phi_2^0)}\\
&=&\frac{\pi(\lambda,\phi_2=\phi_2^0|\phi_1=\phi_1^0)}{\pi(\phi
_2=\phi
_2^0|\phi_1=\phi_1^0)}\\
&=&\pi(\lambda|\phi_1=\phi_1^0,\phi_2=\phi_2^0),
\end{eqnarray*}
which establishes the result.

Finally consider the JC procedure. We have
%
\begin{eqnarray}\label{eq:JCM}
\qquad\pi^{**}_{\cM}(\lambda) \propto
\pi(\lambda|\phi_1=\phi_1^0,\phi_2=\phi_2^0)
\frac{j_{\cM^{**}}(\lambda)}{j(\lambda,\phi_1^0,\phi_2^0)}.
\end{eqnarray}
On the other hand
%
\begin{eqnarray}\label{eq:JCM*}
\pi^{**}_{\cM^*}(\lambda)\propto
\pi^*_{\cM}(\lambda|\phi_2=\phi_2^0)
\frac{j_{\cM^{**}}(\lambda)}{j_{\cM^*}(\lambda,\phi_2^0)},
\end{eqnarray}
where $\pi^*_{\cM}(\lambda|\phi_2=\phi_2^0)$ is proportional to
the JC prior under the $\cM^*$ model, evaluated at
$(\phi_2=\phi_2^0)$, namely
$\pi^*_{\cM}(\lambda,\phi_2=\phi_2^0)$, where
\begin{eqnarray*}
\pi^{*}_{\cM}(\lambda,\phi_2)\propto
\pi(\lambda,\phi_2|\phi_1=\phi_1^0)\frac{j_{\cM^{*}}(\lambda,
\phi_2)}{j(\lambda, \phi_1^0,\phi_2^0)}.
\end{eqnarray*}
Substituting into
(\ref{eq:JCM*}), one obtains (\ref{eq:JCM}).
\end{pf*}

\setcounter{thm}{0}
\begin{lem}\label{lemmaA1}
Assume $(\bet, \sdue) \sim\mathrm{NIGa}(b,g (X^T\cdot X)^{-1},d,a)$ and set
$R_k(\bet, \sdue)=(1+Q_k(\bet)/\sdue)^{-1}$, with
$Q_k(\bet)=\bet^TX^T(I-P_k)X\beta/n$. Then, given $\sdue$,
$Q_k(\bet)/\sdue\sim(g/n) \chi^2_{p-k}(\delta)$, with $\delta=n
Q_k(b)/(g\sdue)$, where $\chi^2_{p-k}(\delta)$ is a chi-squared
distribution with $(p-k)$ degrees of freedom and noncentrality
parameter $\delta$. As a consequence
%
\begin{eqnarray}\label{eq:ER-1}
&&E[R_k(\bet,\sdue)^{-1}]\nonumber\\[-8pt]\\[-8pt]
&&\quad =  \biggl(1+\frac{g}{n}(p-k)+Q_k(b)\frac{d}{a}
\biggr),\nonumber\\
\label{eq:VarR-1}
&&\operatorname{Var}[R_k(\bet, \sdue)^{-1}] \nonumber\\[-8pt]\\[-8pt]
&&\quad=
\frac{2d}{a}Q_k(b) \biggl[\frac{Q_k(b)}{a}+\frac{2g}{n} \biggr] +
\frac{2g^2}{n^2}(p-k).\nonumber
\end{eqnarray}
Furthermore
\begin{longlist}[(ii)]
\item[(i)]if $\bnk=0$, then $R_k(\bet, \sdue)=[1+(g/n)W]^{-1}$ with
$W$ distributed as a (central)
$\chi^2_{p-k}$, whence
%
\begin{eqnarray}\label{eq:ER}
E[R_k(\bet,\sdue)]
&=& \biggl(\frac{2g}{n} \biggr)^{-(p-k)/2}\exp\biggl(\frac{n}{2g} \biggr)\nonumber\\[-8pt]\\[-8pt]
&&{}\cdot\Gamma
\biggl(1-\frac{p-k}{2};\frac{n}{2g} \biggr),\nonumber
\end{eqnarray}
where $\Gamma(\alpha,z)=\int_z^{\infty} \exp(-t) t^{\alpha-1}
\,dt$ is the incomplete gamma function.

\item[(ii)]
If $b_{\setminus k}\neq0$, then the first-order approximation of
$E[R_k(\bet, \sdue)]$ given by the delta method is
%
\begin{eqnarray}
\label{eq:Rapprox}
E[R_k(\bet, \sdue)]&\approx&
\frac{1}{E[R_k(\bet, \sdue)^{-1}]}\nonumber\\[-8pt]\\[-8pt]
&=& \biggl[1+\frac{g}{n}(p-k)+Q_k(b)\frac{d}{a} \biggr]^{-1}.\nonumber
\end{eqnarray}
\end{longlist}
\end{lem}

\begin{pf}
First of all notice that because $X=\break[X_k\vdots X_{\setminus k}]$
and $\bet=[\bet_k^T \vdots\bet_{\setminus k}^T]^T$, we have
$Q_k(\bet)=\break \bet^TX^T M_kX\beta/n= \bet_{\setminus
k}^TX_{\setminus k}^T M_k X_{\setminus k}\bet_{\setminus k}/n$.
Now $\bet_{\setminus k}|\sdue$ is distributed according to a
N$(b_{\setminus k}, g \sdue\Sigma_{\setminus k})$ with
$\Sigma_{\setminus k}=(X_{\setminus k}^T M_k X_{\setminus
k})^{-1}$ (see Searle, \citeyear{Sea1982}, Section 10.5), and consequently $(n/g)
Q_k(\bet)/\sdue$ given $\sdue$
is distributed according to a $\chi^2_{p-k}(\delta)$
distribution, where ${p-k}$ are the degrees of freedom and
$\delta= (n/g)\* Q_k(b)/\sdue$ is the noncentrality parameter (see
Muirhead, \citeyear{Mui1982}, page 26). Now recalling that the expected value and
variance
of a $\chi^2_{p-k}(\delta)$ distribution are respectively
$p-k+\delta$ and $2(p-k)+4\delta$, (\ref{eq:ER-1})
follows immediately from
$E[R_k(\bet,\sdue)^{-1}]=E^{\sdue}\{1+E^{\bet|\sdue}[Q_k(\bet
)/\sdue]\}$,
and $E(1/\sdue)=d/a$.

Similarly (\ref{eq:VarR-1}) follows from
\begin{eqnarray*}
\operatorname{Var}[R_k(\bet, \sdue)^{-1}]&=& \operatorname
{Var}[Q_k(\bet)/\sdue]\\
&=&\operatorname{Var}^{\sdue} \biggl[ E^{\bet|\sdue} \biggl(\frac{Q_k(\bet
)}{\sdue} \biggr) \biggr]\\
&&{}+
E^{\sdue} \biggl[ \operatorname{Var}^{\bet|\sdue} \biggl(\frac{Q_k(\bet
)}{\sdue} \biggr) \biggr]\\
&=&\frac{g^2}{n^2} \biggl\{\operatorname{Var}^{\sdue} \biggl[p-k+\frac{n}{g
\sdue}Q_k(b) \biggr]\\
&&{}+ E^{\sdue} \biggl[2(p-k)+4\frac{nQ_k(b)}{g
\sdue} \biggr] \biggr\}
\end{eqnarray*}
and $\operatorname{Var}(1/\sdue)=2d/a^2$.

\begin{longlist}[(ii)]
\item[(i)] If $b_{\setminus k}=0$, then $\delta=0$, so that
$W=(n/g)Q_k(\bet)/\break \sdue$ is distributed as a (central)
$\chi^2_{p-k}$. Thus
$E[R_k(\bet,\break \sdue)]=E^{W} [ (1+(g/n)W )^{-1}
]$
whose analytical expression is given in
(\ref{eq:ER}).

\item[(ii)]
If $b_{\setminus k}\neq0$, writing $E[R_k(\bet,
\sdue)]=E[1/R_k(\bet,\break \sdue)^{-1}]$ and recalling that the
first-order approximation gives $E(1/W) \approx1/(E(W))$ for an
arbitrary random variable $W$, we obtain (\ref{eq:Rapprox}).\qed
\end{longlist}
\noqed
\end{pf}

\begin{pf*}{Proof of Proposition \ref{prop:kl}}
The NIGa$(b_k,\break g_k (X_k^TX_k)^{-1},d_k,a_k)$ distribution on
$(\betk^*, \sduek)$ can be written as
\begin{eqnarray*}
\pi(\betk^*, \sduek)
&\propto&\exp\biggl\{-\frac{1}{2\sduek}a+b^TX_k^T X_k \frac{\betk^*
}{g\sduek}\\
&&\phantom{\exp\biggl\{}{} -\frac{1}{2g}\frac{\betk^{*T} X_k^T X_k
\betk^*}{\sduek}\\
&&\hspace*{5pt}\phantom{\exp\biggl\{}{}
+\frac{d+p+2}{2}\log\biggl(\frac{1}{\sduek} \biggr) \biggr\},
\end{eqnarray*}
thus
it belongs to the exponential family with ``canonical
statistics''
given by $1/\sduek$, $\betk^*/\sduek$, $\betk^{*T} X_k^T X_k
\betk^*/\sduek$ and $\log(1/\sduek)$. Applying Theorem 1 of
Consonni, Guti\'errez-Pe\~na and
Veronese (\citeyear{ConGutVer2007}), it follows that the KL-divergence
between $\pi^{\mathrm{KL}}$ and a NIGa$(b_k,g_k (X_k^T\cdot X_k)^{-1}, d_k,a_k)$
distribution is minimized for values $\bkl, \gkl, \dkl$ and $\akl$
which are a solution of the following system:
%
\begin{eqnarray}
\label{eq:S1}
&&E^{\mathrm{KL}} (1/{\sduek} )\nonumber\\[-8pt]\\[-8pt]
&&\quad =E^{\mathrm{NIGa}} (1/{\sduek} ),\nonumber
\\
\label{eq:S2}
&&E^{\mathrm{KL}} (\betk^*/{\sduek} )\nonumber\\[-8pt]\\[-8pt]
&&\quad=E^{\mathrm{NIGa}} (\betk^*/{\sduek} ),\nonumber
\\
\label{eq:S3}
&&E^{\mathrm{KL}} (\betk^{*T} X_k^T X_k
\betk^*/\sduek)\nonumber\\[-8pt]\\[-8pt]
&&\quad=E^{\mathrm{NIGa}} (\betk^{*T}X_k^T X_k
\betk^*/\sduek),\nonumber\\
\label{eq:S4}
&&E^{\mathrm{KL}} (\log(1/{\sduek}) )\nonumber\\[-8pt]\\[-8pt]
&&\quad=E^{\mathrm{NIGa}} (\log(1/{\sduek}) ),\nonumber
\end{eqnarray}
where $E^{\mathrm{KL}}$ denotes expectation w.r.t. the
KL-projec\-tion prior induced by the  NIGa$(\bet, \sdue; b,g
(X^TX)^{-1},\break d,a)$, while $E^{\mathrm{NIGa}}$ denotes expectation w.r.t. the\break
NIGa$(\betk^*,
\sduek; \bkl,\gkl(X_k^TX_k)^{-1}, \dkl,\akl).$ Recalling
(\ref{eq:KL-projection-betak}) and (\ref{eq:KL-projection-sduek}),
that is,
$ \betk^{\perp}=(X^T_k X_k)^{-1}X_k^TX \beta$, $ \sigma_k^{2
\perp}=
\sigma^2+ Q_k(\bet)$,
we can compute the terms involving $E^{\mathrm{KL}}$ in
the previous equations substituting $(\betk^*, \sduek)$ with the
corresponding expression of $(\betk^{\perp}, \sigma_k^{2 \perp})$
and using the prior $\pi(\bet,\sdue)$.

First of all recall that if $Y$ is a normal vector with
variance matrix $\Sigma$, then $Y^TAY$ and $CY$ are
stochastically independent if and only if $C \Sigma A=0$; similarly
$Y^TAY$ and $Y^TDY$ are
stochastically independent if and only if $A \Sigma D=0$ (with $A$,
$C$ and $D$ being suitable matrices).
It follows that under $\pi$ and given $\sdue$, $Q_k(\bet)$ and
$\betk^{\perp}$ as well as $Q_k(\bet)$ and
${\betkl}^T X_k^T X_k \betkl$ are independent; the latter implies
that also
${\betkl}^T X_k^T X_k \betkl$ and $\sduekl$ are independent, given
$\sdue$.
The proof is a straightforward calculation.

Consider now  (\ref{eq:S2}). The left-hand side is
equal to
\begin{eqnarray*}
E \biggl(\frac{\betkl}{\sduekl} \biggr)&=&E^{\sdue} \biggl\{
E^{\bet|\sdue} \biggl[\frac{1}{\sdue+Q_k(\bet)} \biggr]\\
&&\phantom{E^{\sdue} \biggl\{}
{}\cdot E^{\bet|\sdue} [(X^T_k X_k)^{-1}X_k^T X \beta] \biggr\}\\
&=&(X^T_k X_k)^{-1}X_k^TX b E (1/\sduek),
\end{eqnarray*}
while the right-hand side is equal to
$ E^{\mathrm{NIGa}}(\betkst/\sduek)=\bkl E^{\mathrm{NIGa}}(1/\sduek)$.
Using  (\ref{eq:S1}) it follows that
%
\begin{eqnarray}\label{eq:bkl}
\bkl=(X^T_k X_k)^{-1}X_k^TX b.
\end{eqnarray}

Consider (\ref{eq:S3}). First of all notice that, using (\ref
{eq:KL-projection-betak}),
${\betkl}^T X^T X \betkl=\bet^T X^T P_k X \bet$.
Thus the left-hand side can be written, recalling the independence of
$\sduekl$ and
${\betkl}^T X^T X \betkl$, given $\sdue$, as
\begin{eqnarray*}
&&E^{\sdue}[E^{\bet|\sdue}(1/\sduekl)E^{\bet|\sdue}(\bet^T
X^T P_k X \bet)]\\
&&\quad=
E^{\sdue}\bigl\{E^{\bet|\sdue}\bigl[(1/\sduekl)\\
&&\qquad\phantom{E^{\sdue}\{E^{\bet|\sdue}[}
{}\cdot\bigl(\operatorname{tr}(\sdue g P_k P) + b^T X^T
P_k X
b\bigr)\bigr]\bigr\}\\
&&\quad
= E^{\sdue}[(kg \sdue+ b^T X^T P_k X b) E^{\bet|\sdue}(1/\sduekl)]\\
&&\quad =
kg E[R_k(\bet,\sdue)] + b^T X^T P_k X b E(1/\sduekl),
\end{eqnarray*}
where $R_k(\bet,\sdue)=[1+Q_k(\bet)/\sdue]^{-1}$.

The right-hand side is equal to
\begin{eqnarray*}
& & E^{\sduek}[(1/\sduek)E^{\betkst|\sduek}({\betkst}^T X_k^T X_k
\betkst)]\\
&&\quad= E^{\sduek}\{(1/\sduek)[\operatorname{tr}(\sduek\gkl(X_k^T X_k)^{-1}(X_k^T
X_k))\\
&&\qquad\phantom{E^{\sduek}\{(1/\sduek)[}\hspace*{38pt}
{}+{\bkl}^T(X_k^T X_k)\bkl]\}\\
&&\quad= k \gkl+b^TX^TP_kXb E^{\sduek}(1/\sduek),
\end{eqnarray*}
substituting
the expression of $\bkl$ given in (\ref{eq:bkl}). Equating the
left- and right-hand sides and using (\ref{eq:S1}) we
obtain
%
\begin{eqnarray}\label{eq:gkl} \gkl=gE[R_k(\bet, \sdue)].
\end{eqnarray}

Consider (\ref{eq:S4}). The left-hand side can be written
as $ E[\log(1/\sdue)]+E[\log(R_k(\bet,\sdue))]$ with
$E[\log(1/\break\sdue)]=\Psi(d/2)-\log(a/2),$ where
$\Psi(\alpha)=\frac{\partial}{\partial\alpha}\cdot\break
\log(\Gamma(\alpha))$ is the digamma function. The right-hand side
is equal to $\Psi(\dkl/2)-\log(\akl/2)$ and thus we obtain
%
\begin{eqnarray}\label{eq:Psi}
&&\Psi(d/2)-\log(a/2)+E[\log(R_k(\bet,\sdue)]\nonumber\\[-8pt]\\[-8pt]
&&\quad=\Psi(\dkl/2)-\log(\akl/2).\nonumber
\end{eqnarray}

Assume now that $b_{\setminus k}=0$ and consider last
(\ref{eq:S1}). First notice that the left-hand side can be
written as $E[R_k(\bet, \sdue)/\sdue]$, while the right-hand side
is equal to $\dkl/\akl$.
Since, from Lemma \ref{lemmaA1},
$R_k(\bet, \sdue)$ is independent of $\sdue$ when $b_{\setminus
k}=0$, (\ref{eq:S1}) becomes
%
\begin{eqnarray}\label{eq:S1bis}
E(1/\sdue)E[R_k(\bet,\sdue)]=\dkl/\akl,
\end{eqnarray}
which implies
%
\begin{eqnarray}
\label{eq:akl} \akl= \dkl\frac{a}{d}
\frac{1}{E[R_k(\bet,\sdue)]}.
\end{eqnarray}

Substituting (\ref{eq:akl}) into (\ref{eq:Psi}) we obtain
%
\begin{eqnarray}\label{eq:Psibis}
&&\Psi(\dkl/2)-\log(\dkl/2)\nonumber\\
&&\quad=\Psi(d/2)-\log(d/2)+E\{\log[R_k(\bet
,\sdue)]\}\\
&&\qquad{}-\log\{E[R_k(\bet,\sdue)]\}.\nonumber
\end{eqnarray}

Consider now the case $b_{\setminus k}\neq0$. In order to obtain
an explicit expression of (\ref{eq:gkl}), we use the approximation
of $E[R_k(\bet,\sdue)]$ given in (\ref{eq:Rapprox}), so that
%
\begin{eqnarray}\label{eq:gklapp}
\gkl&\approx&\frac{g}{E[R^{-1}_k(\bet,
\sdue)]}\nonumber\\[-8pt]\\[-8pt]
&=&\frac{g}{ [1+g/n(p-k)+Q_k(b)d/a ]}.\nonumber
\end{eqnarray}

Furthermore, we can
still use (\ref{eq:S1bis}) as an approximation of
(\ref{eq:S1}) to the first order.
Thus we have
%
\begin{eqnarray}\label{eq:aklapp}
\akl&\approx& \dkl\frac{a}{d} \frac{1}{E[R_k(\bet,\sdue)]}\nonumber\\
&\approx&
\dkl\frac{a}{d} E[R^{-1}_k(\bet,\sdue)]\\
&=& \dkl\frac{a}{d}
\biggl[1+\frac{g}{n}(p-k)+Q_k(b)\frac{d}{a} \biggr], \nonumber
\end{eqnarray}
using (\ref{eq:Rapprox}).

Using the first approximation of (\ref{eq:aklapp}), formula
(\ref{eq:Psibis}) still holds in an approximate way.

Finally (\ref{eq:Psibis})
reduces to
\begin{eqnarray*}\label{eq:psiapp}
&&\Psi(\dkl/2)-\log(\dkl/2)\\
&&\quad\approx \Psi(d/2)-\log(d/2) - \frac{1}{2}\frac{\operatorname
{Var}([R_k(\bet,
\sdue)])}{E[R_k(\bet, \sdue)]^2},
\end{eqnarray*}
using the further second-order approximation\break $E[\log(U)]\approx\log
[E(U)]-(1/2)\operatorname{Var}(U)/[E(U)]^2$,
for a positive random variable $U$.

Since $\operatorname{Var}(U)=\operatorname{Var}(1/U^{-1})\approx
[1/E(U^{-1})]^4 \cdot\break\operatorname{Var}(U^{-1})$ and
$E(U)=E(1/U^{-1}) \approx1/E(U^{-1})$ we conclude
\begin{eqnarray*}
&&\Psi(\dkl/2)-\log(\dkl/2)\\
&&\quad \approx \Psi(d/2)-\log(d/2) -
\frac{1}{2}\frac{\operatorname{Var}([R^{-1}_k(\bet, \sdue
)])}{E[R^{-1}_k(\bet,
\sdue)]^2}
\end{eqnarray*}
with $E[R^{-1}_k\bet, \sdue)]$ and $\operatorname{Var}[R^{-1}_k(\bet
, \sdue)]$
given in
(\ref{eq:ER-1}) and (\ref{eq:VarR-1}).
\end{pf*}
\end{appendix}

\section*{Acknowledgments}

GC's research was supported in part by MIUR, Rome (PRIN 2003138887 and PRIN
2005132307) and the
University of Pavia. PV's research was supported in part by
MIUR, Rome (PRIN 2003138887 and PRIN 2005132307) and by L.~Bocconi
University. Part of this
work was written while the authors visited Universit\'{e} Paris-Sud,
Orsay, France. They are
grateful to Gilles Celeux, Jean Michel Marin and Christian Robert for
helpful discussions.
Useful comments on some of the ideas presented in this paper were also
provided by Eduardo
Guti\'{e}rrez-Pe\~{n}a and Manuel Mendoza. The support of the
Executive Editor and the reviewers' comments are gratefully
acknowledged.

\end{document}